\documentclass[aps,compressed, 12pt, preprint, preprintnumbers, amsmath,amssymb,nofootinbib,superscriptaddress,hyperref]{revtex4-1} 
\usepackage[usenames,dvipsnames]{color}  \usepackage{graphicx}
\usepackage{slashed}
\usepackage{amsmath,amssymb,color,colortbl}
\usepackage[right=25truemm,left=25truemm,top=36truemm,bottom=30truemm]{geometry}
\usepackage{fancyhdr}
\usepackage{here}
\usepackage[subrefformat=parens]{subcaption}
\usepackage[samesize]{cancel} 
\usepackage{comment}
\usepackage{caption}
\usepackage{subcaption}
\usepackage[colorlinks=true,citecolor=darkred,urlcolor=darkred, pdfborder={0 0 0}]{hyperref}
\usepackage[normalem]{ulem}
\definecolor{darkred}{rgb}{0.6,0,0}
\newcommand{\beq}{\begin{equation}}
\newcommand{\eeq}{\end{equation}}
\newcommand{\bea}{\begin{eqnarray}}
\newcommand{\eea}{\end{eqnarray}}
\def\beq#1\eeq{\begin{equation}#1\end{equation}}
\def\bal#1\eal{\begin{align}#1\end{align}}

\newcommand{\di}[1]{\overline{#1}}

\newcommand{\beas}{\begin{eqnarray*}}
\newcommand{\eeas}{\end{eqnarray*}}
\definecolor{linkcolor}{rgb}{0,0,0.5}
\begin{document}
\begin{flushright}
\preprint{\textbf{OU-HET-1078}}
\end{flushright} 
\bibliographystyle{unsrt} 
%%%%%%%%%%%%%%%%%%%%%%%%%%%%%%%%%%%%%%%%%%%%%%%%%%%%%%%
\title{Alternative dark matter phenomenology in a general $U(1)_X$ extension of the Standard Model}
%%%%%%%%%%%%%%%%%%%%%%%%%%%%%%%%%%%%%%%%%%%%%%%%%%%%%%%%
\author{Arindam Das}\email{arindam.das@het.phys.sci.osaka-u.ac.jp}
\affiliation{Department of Physics, Osaka University, Toyonaka, Osaka 560-0043, Japan}
\author{Kazuki Enomoto}\email{kenomoto@het.phys.sci.osaka-u.ac.jp}
\affiliation{Department of Physics, Osaka University, Toyonaka, Osaka 560-0043, Japan}
\author{Shinya Kanemura}\email{kanemu@het.phys.sci.osaka-u.ac.jp}
\affiliation{Department of Physics, Osaka University, Toyonaka, Osaka 560-0043, Japan}
%%%%%%%%%%%%%%%%%%%%%%%%%%%%%%%%%%%%%%%%%%%%%
\begin{abstract}
The existence of the neutrino mass and flavor mixing have been experimentally verified. 
These phenomena strongly motivate to extend the Standard Model (SM). 
Amongst many possibilities, a simple and interesting extension of the SM can be investigated using a general 
U$(1)_X$ extension of the SM gauge group. Demanding the cancellation of the gauge and mixed gauge gravity anomalies, 
three right handed neutrinos are introduced in this model where the U$(1)_X$ charge assignment becomes a linear combination of U$(1)_{\rm{B-L}}$ and U$(1)_Y$ hyper-charges. 
After the U$(1)_X$ breaking, an additional neutral gauge boson, $Z^\prime$ is evolved and the neutrino mass is generated by the seesaw mechanism. 
In such a model we investigate the properties of a Dark Matter (DM) candidate which is a massive weakly interacting particle and Dirac type in nature. The stability of the DM is protected by its U$(1)_X$ charge. Using the current bounds on the search results of $Z^\prime$ at the Large Hadron Collider (LHC) and the dark matter relic abundance we find a phenomenologically viable parameter space of our scenario. 
\end{abstract}
%%%%%%%%%%%%%%%%%%%%
\maketitle
%%%%%%%%%%%%%%%%%%%%%%%%%%%%%%%%%%%%
%\section{Introduction}
%\label{intro}
%%%%%%%%%%%%%%%%%%%%%%%%%%%%%%%%%%%%%
The existence of the tiny neutrino mass and flavor mixing give clear indications to beyond the SM physics \cite{Zyla:2020zbs}. A large variety of scenarios have been predicted to explain the origin of the tiny neutrino mass and the flavor mixing. Probably the most simple scenario could be considered as the extension of the SM by an SM-singlet heavy Majorana neutrino which generates the tiny neutrino mass term by the seesaw mechanism \cite{Minkowski:1977sc,Mohapatra:1979ia,Schechter:1980gr,Yanagida:1979as,GellMann:1980vs,Glashow:1979nm,Mohapatra:1986aw,Mohapatra:1986bd}. Another interesting approach is to extend the SM by an additional U$(1)$ gauge group where three generations of the right handed neutrinos are automatically introduced to cancel the gauge and mixed gauge-gravity anomalies. A very well motivated example is the B$-$L (baryon minus lepton) scenario \cite{Marshak:1979fm,Mohapatra:1980qe,Wetterich:1981bx,Masiero:1982fi,Buchmuller:1991ce} where three generations of the SM-singlet right handed neutrinos are introduced. In the B$-$L model there is an SM-singlet scalar which couples to the right handed neutrinos conserving the B$-$L symmetry. After the B$-$L symmetry breaking, the Majorana mass of the right handed neutrinos are generated which further allows the seesaw mechanism to generate the tiny light neutrino masses. The introduction of the additional U$(1)_{\rm{B-L}}$ symmetry in the SM also explores an interesting BSM feature which introduces a neutral gauge boson $Z^\prime$- acquiring mass after the B$-$L symmetry breaking- being studied for a long period of time for interesting phenomenological aspects \cite{DelAguila:1995fa,Leike:1998wr,Barger:2003zh,Rizzo:2006nw,Montero:2007cd,Langacker:2008yv}. 

Another interesting scenario is a gauged U$(1)_X$ extension of the SM where the U$(1)_X$ charges of the SM fields are given by a linear combination of the U$(1)_Y$ and the U$(1)_{\rm{B-L}}$. As both of the SM and the U$(1)_{\rm{B-L}}$ are the anomaly free scenarios, the linear combination of these two scenarios will be free from all the gauge and mixed gauge-gravity anomalies. The U$(1)_X$ symmetry is broken by the vacuum expectation value (VEV) of the SM-singlet U$(1)_X$ scalar. There is a BSM neutral gauge boson $Z^\prime$ which acquires mass after the U$(1)_X$ symmetry breaking. At the same time the U$(1)_X$ scalar generates Majorana mass term for the SM-singlet right handed neutrinos. This Majorana mass term induces the seesaw mechanism in the U$(1)_X$ case to produce the light neutrino masses \cite{Appelquist:2002mw,Das:2017flq,Choudhury:2020cpm}. 

The observations of the cosmic microwave background (CMB) anisotropy by the Wilkinson Microwave Anisotropy Probe (WMAP) \cite{Hinshaw:2012aka} and the PLANCK satellite \cite{Aghanim:2018eyx} determine the energy budget of the present universe where $73\%$ is the dark energy, $23\%$ is dark matter and only $4\%$ is the baryonic matter. Since there is no suitable DM candidate in the SM, we need to extend it. The introduction of a Weakly Interacting Massive Particle (WIMP) \cite{Lee:1977ua} as a suitable DM candidate is one of the interesting and promising approaches which has been widely studied \cite{Arcadi:2017kky,Okada:2018ktp}. The WIMP DM candidate interacts with the SM particles and it was in thermal equilibrium in the early universe. 
 
To introduce a DM candidate in this model we need to make further extension leading to a more complete scenario\cite{Okada:2010wd,Kanemura:2011vm,Bandyopadhyay:2011qm, Okada:2016gsh,Okada:2016tci,Bandyopadhyay:2017bgh,Das:2019pua,Kang:2019sab,Bonilla:2018ynb}. Here we introduce a cold DM candidate charged under the U$(1)_X$ in such a way that it does not interact with the SM candidates. 
Due to the non-zero U$(1)_X$ charge of the DM candidate interacts with the $Z^\prime$ boson. 
Such a scenario has been studied in \cite{FileviezPerez:2018toq,Duerr:2015wfa} under the U$(1)_{\rm{B-L}}$ scenario. 
In the general U$(1)_X$ model we can write the U$(1)_X$ charges of the SM particles as the linear combination of the SM U$(1)_Y$ hyper-charge and the U$(1)_{\rm{B-L}}$ charges. In the recent years the LHC is searching for the $Z^\prime$ bosons. In the Run-2 at $\sqrt{s}=13$ TeV, the ATLAS \cite{Aad:2019fac} collaboration was searching for the 
evidence of the heavy resonance from the dilepton signal under the Sequential Standard Model (SSM) scenario where the coupling between the $Z^\prime$ boson and the SM fermions are same as those in the SM. The neutral gauge boson is heavier than the $Z$ boson. Comparing the dilepton results in the U$(1)_X$ case with the LHC results we find the bounds on the U$(1)_X$ coupling and the $Z^\prime$ mass. Finally we use the allowed parameter space for the DM study.

We identify the U$(1)_X$ gauge group as the linear combination of the SM U$(1)_{\rm{Y}}$ and the U$(1)_{\rm{B-L}}$ gauge group.
The U$(1)_X$ charges of the particles of the model are dependent on $x_H$ and $x_\Phi$ which are the real parameters \cite{Oda:2015gna,Das:2016zue} and 
the particle content is given in Tab.~\ref{tab1}.  We can express the U$(1)_X$ charge $(Q_X)$ as $Q_X= x_H Q_{\rm{Y}}+ x_\Phi Q_{\rm{B-L}}$ where $Q_Y$ is the SM U$(1)_{\rm{Y}}$ hyper-change and $Q_{\rm{B-L}}$ is the B$-$L charge. In this case the charge $x_\Phi$ is not an independent parameter. It appears as a product with the gauge coupling of U$(1)_X$ $(g^\prime)$. If we fix $x_\Phi=1$ in the rest of the analysis. Hence $x_H$ can be considered as the `angle' between the U$(1)_{\rm{Y}}$ and U$(1)_{\rm{B-L}}$ directions. We can reproduce the B$-$L model considering $x_H =0$. The limits $x_H \to \infty$ and $x_H \to -\infty$ U$(1)_X$ is respectively aligned and anti-aligned to the U$(1)_Y$ direction.
%%%%%%%%%%%%%%%%%%%%%%%%%%%%
\begin{table}[H]
\begin{center}
\small
\begin{tabular}{|c|c|c|c|c|c|c||c|c|} \hline
 & $q_{L_{i}}$ & $u_{R_{i}}$ & $d_{R_{i}}$ & $\ell_{L_{i}}$ & $e_{R_{i}}$ & $N_{R_{\alpha}}$ & $H$ & $ \Phi $ \\ \hline\hline
SU$(3)_{\rm{C}}$ & ${\bf 3}$ & ${\bf 3}$ & ${\bf 3}$ & ${\bf 1}$ & ${\bf 1}$ & ${\bf 1}$ & ${\bf 1}$ & ${\bf 1}$ \\ \hline \hline
SU$(2)_{\rm{L}}$ & ${\bf 2}$ & ${\bf 1}$ & ${\bf 1}$ & ${\bf 2}$ & ${\bf 1}$ & ${\bf 1}$ & ${\bf 2}$ & ${\bf 1}$ \\ \hline \hline 
U$(1)_{\rm{Y}}$ & $1/6$ & $2/3$ & $-1/3$ & $-1/2$ & $-1$ & $0$ & $1/2$ & $0$ \\ \hline \hline
U$(1)_X$ & $\frac{1}{6}x_H + \frac{1}{3}x_\Phi$ & $\frac{2}{3}x_H+\frac{1}{3}x_\Phi$ & $-\frac{1}{3}x_H+\frac{1}{3}x_\Phi$ & $-\frac{1}{2}x_H-x_\Phi$ & $-x_H-x_\Phi$ & $-x_\Phi$ & $\frac{1}{2}x_H$ & $2x_\Phi$ \\ \hline
\end{tabular}
\caption{Particle content of  the minimal U(1)$_X$ model. In addition to the SM particle content, three RHNs ($N_{R_\alpha}$) and 
  a SM-singlet Higgs field ($\Phi$) are introduced. Here $i(=1,~2,~3)$ is the family index, and $x_H$, $x_\Phi$ are real parameters. Since the U(1)$_X$ gauge coupling is a free parameter in the model, we fix $x_\Phi=1$ without loss of generality. }
\label{tab1}
\end{center}
\end{table}
%%%%%%%%%%%%%%%%%%%%%

This scenario is free from all the gauge and gravitational anomalies due to the three generations of the right handed neutrinos. 
The three right handed neutrinos are involved in the seesaw mechanism to generate the light neutrino mass.
The relevant for the Lagrangian manifesting the light neutrino mass can be written as 
\bea
\mathcal{L}_{\rm{Yukawa}}\supset -Y^{ij} \overline{\ell_L^i} \tilde{H} N_R^j -\frac{1}{2} Y_N^\alpha \overline{{N_R^\alpha}^C} N_R^\alpha \Phi+h.c.
\label{Majo}
\eea
After the U$(1)_X$ breaking, we obtain the Majorana mass term from Eq.~(\ref{Majo}), $M_N^\alpha=\frac{Y_N^\alpha v_\Phi}{\sqrt{2}}$
where $<\Phi>=\frac{v_\Phi}{\sqrt{2}}$. The first term of Eq.~(\ref{Majo}) generates the Dirac mass term after the electroweak symmetry breaking, $m_{D}^{ij}=\frac{Y^{ij} v_H}{\sqrt{2}}$, 
where $<H>=\frac{v_H}{\sqrt{2}}$.
Hence the induced seesaw mechanism can explain the origin of the tiny neutrino masses and the flavor mixings.
The indices $i,j$ and $\alpha$ run over three generations. The heavy neutrino phenomenologies of this model have been studied in \cite{Das:2019fee,Chiang:2019ajm}.

The renormalizable scalar potential for $H$ and $\Phi$ can be written as 
\bea  
V= \lambda \Big(H^\dagger H -\frac{v_H^2}{2}\Big)^2 + \tilde{\lambda} \Big(\Phi^\dagger \Phi -\frac{v_\Phi^2}{2}\Big)^2+\lambda^\prime \Big(H^\dagger H -\frac{v_H^2}{2}\Big) \Big(\Phi^\dagger \Phi -\frac{v_\Phi^2}{2}\Big)
\label{scl}
\eea
Here $\lambda$, $\tilde{\lambda}$ and $\lambda^\prime$ are the real couplings with $v_H=246$ GeV and $v_\Phi$ is the free parameter. 
Using the stationary conditions of the potential with respect to the scalar fields at their VEVs. 
One can calculate the mass matrix for the scalar fields as 
\bea
M = 
\begin{pmatrix}
2\lambda v_H^2& \lambda^\prime v_H v_\Phi\\
\lambda^\prime v_H v_\Phi&2\tilde{\lambda} v_\Phi^2
\end{pmatrix}.
\eea
The scalar are fields expanded around the VEVs such that $H^T=\frac{1}{\sqrt{2}}(v+h_1^\prime~~0)$ and $\Phi=\frac{1}{\sqrt{2}} (v_\Phi+h_2^\prime)$
where the states $h_1^\prime$ and $h_2^\prime$  can be mixed using a $2\times2$ orthogonal matrix and the mixing angle can be written as $\beta$.
We have $\tan2\beta=-\frac{\lambda^\prime v_H v_\Phi}{\lambda v_H^2 -\tilde{\lambda} v_\Phi^2}$ which can be constrained from \cite{Khachatryan:2016whc,Jana:2018rdf}.
The condition for the scalar potential for bounded from below could be written as $|\lambda^\prime| < 2 \sqrt{\lambda \tilde{\lambda}}$ \cite{Oda:2015gna,Das:2016zue}.
From the kinetic terms of the scalar fields and expanding the covariant derivatives the mass term for the $Z^\prime$ boson and its mass is 
$M_{Z^\prime} = g_X \sqrt{4 x_\phi^2 v_\Phi^2+ x_H^2 \frac{v_H^2}{4}} $ and this reduces to $2 g_X v_\Phi$ in the assumption $v_\Phi \gg v_H$.

In this model we introduce a DM candidate $(\chi_L, \chi_R)$ which is SM-singlet and weakly interacting in nature and having U$(1)_X$ charge $n_\chi$.
The fermions $\chi_{L,R}$ interact with the $Z^\prime$ as 
\bea
\mathcal{L}_{\rm{kin}} = i\overline{\chi_L} \gamma_\mu \mathcal{D}^\mu \chi_L - i\overline{\chi_R} \gamma_\mu \mathcal{D}^\mu \chi_R
\label{Xm}
\eea
where $\mathcal{D}^\mu= \partial^\mu + i g_X n_\chi {Z^\mu}^\prime$. To prevent the decays of $\chi_{L, R}$ into the right handed neutrinos and other particles we prohibit $|n_\chi|=\pm x_\Phi, \pm 3 x_\Phi$ which ensures the stability of $\chi_{L, R}$. Therefore we can consider $\chi_{L,R}$ as potential DM candidate. Using the Dirac mass term of $\chi_L$ and $\chi_R$ in Eq.~(\ref{Xm}) we can further write
\bea
\mathcal{L}^{\rm{DM}} = i \overline{\chi} \gamma_\mu \partial^\mu \chi- g_X n_\chi \gamma_\mu {Z^\mu}^\prime \chi + (m_{\rm{DM}} \overline{\chi_L} \chi_R + h. c.)
\label{XDM}
\eea
where $\chi= \chi_L+\chi_R$. In this analysis we consider $x_\Phi=1$.
We consider a scenario which is sensitive to the UV theory. 
Hence the neutrinos and the DM candidate could mix through higher dimensional and non-renormalizable operators for odd $n_\chi$, however, 
the safe choices for $n_\chi$ could be even and fractional numbers. However, the presence of $x_H$ through the interaction between the $Z^\prime$ and the SM particles will make our scenario interesting. In our analysis, for simplicity we neglect the kinetic mixing between the $Z$ and $Z^\prime$. 
Finally we estimate the constraints on the $g_X$ vs $M_{Z^\prime}$ plane comparing the dilepton production cross section with the recent LHC study of the dilepton signature from the heavy resonance \cite{Aad:2019fac} and apply these results to estimate the parameter space allowed by the DM relic abundance $\Omega h^2=0.120\pm0.001$ \cite{Aghanim:2018eyx}.
%%%%%%%%%%%%%%%%%%%%%%%%%%%%%%%%%%%%%%%%%%%%%%
%\section{Constraints on $g_X$ and $M_{Z^\prime}$ from the LHC at $\sqrt{s}=13~\rm{TeV}$} 
%\label{const}
%%%%%%%%%%%%%%%%%%%%%%%%%%%%%%%%%%%%%%%%%%%%

\begin{figure}[h]
\centering
\includegraphics[width=0.485\textwidth]{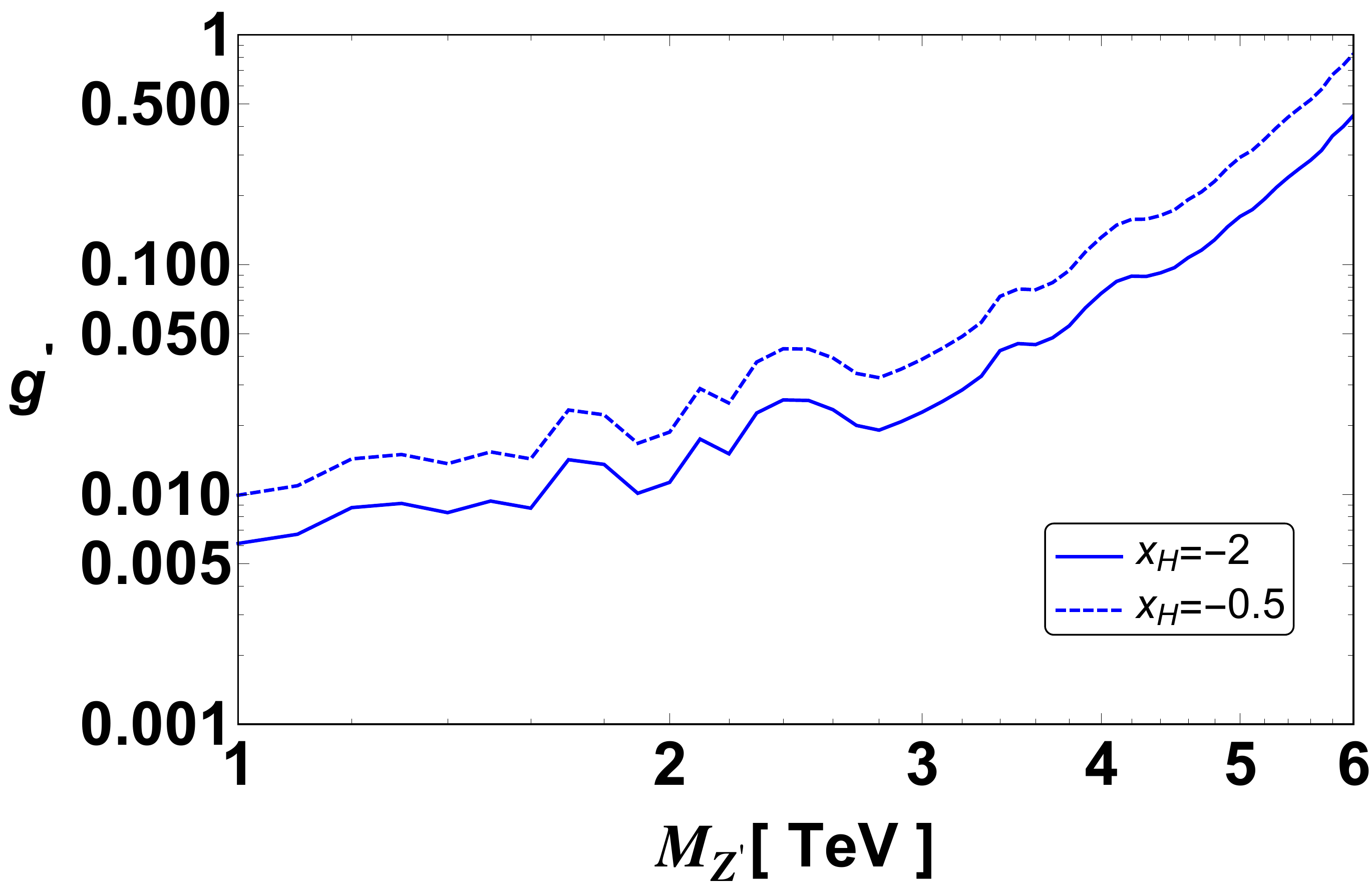} 
\includegraphics[width=0.485\textwidth]{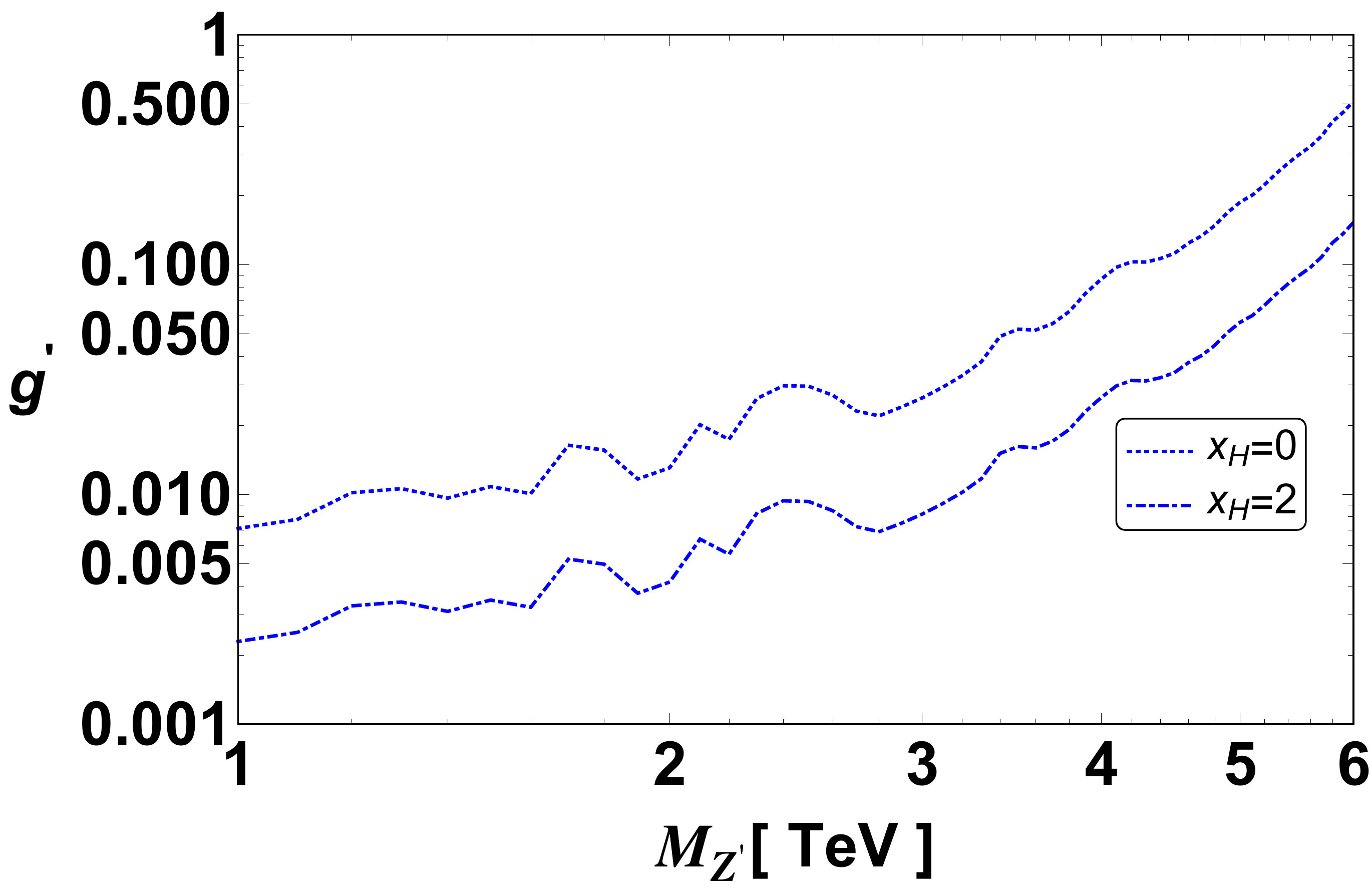} 
\caption{The constraints on the $g_X-M_{Z^\prime}$ plane for different $x_H$ fixing $x_\Phi=1$. 
We show the bounds for $x_H < 0$ in the left panel and those for $x_H \geq 0$ in the right panel.}
\label{bounds1}
\end{figure}

We study the production of $Z^\prime$ at the LHC from the proton proton interaction into a pair of leptons. 
In this case we consider the ATLAS results at the $139$ fb$^{-1}$ luminosity from the Ref. \cite{Aad:2019fac}.
\footnote{At the time analysis we found a conference proceeding from CMS collaboration at the $140$ fb$^{-1}$  \cite{CMS:2019tbu} which is very close to the ATLAS result \cite{Aad:2019fac}. Therefore in this article we use the ATLAS result as the final result will be very close.} 
We calculate the $pp\to Z^\prime \to \ell^+ \ell^-$ process at the $13$ TeV for $\ell= e, \mu$ considering $x_\Phi=1$ and $x_H= -2, -0.5, 0,2$.
We compare our cross section with the ATLAS result \cite{Aad:2019fac} for combined electron and muon final state.
The ATLAS collaboration has considered two different models, namely, SSM and $Z^\prime_\psi$ \cite{Langacker:2008yv} where the $Z^\prime$ boson decays into the electron and muon in pair. We conservative consider this analysis and find the limits on the $g_X-M_{Z^\prime}$ plane for $1$ TeV $\leq M_{Z^\prime} \leq 6$ TeV at the $13$ TeV LHC. Computing the production cross section for the model, $\sigma_{\rm{Model}}(pp\to Z^\prime \to 2e+2\mu)$ with a U$(1)_X$ coupling $g_X^{\rm{Model}}$ and comparing the observed ATLAS bound on the dilepton production cross section $(2e+2\mu)$ for ${\Gamma}{m}=3\%$ in the SSM case $(\sigma_{\rm{ATLAS}}^{\rm{Observed, SSM}})$, we find the bounds on the $g_X$ corresponding to the $Z^\prime$ masses according to
\bea
g_X = \sqrt{\sigma_{\rm{ATLAS}}^{\rm{Observed, SSM}}\ast \Big(\frac{(g_X^{\rm{Model}})^2}{\sigma_{\rm{Model}}(pp\to Z^\prime \to (2e+2\mu))}\Big)}
\label{bounds}
\eea
since the cross section varies with the square of the U$(1)_X$ coupling. 
The corresponding bounds on the $g_X$ and corresponding $M_{Z^\prime}$ values for different $x_H$ can be found in Fig.~\ref{bounds1}.
%%%%%%%%%
%%%%%%%%%%%%%%%%%%%
%\section{Calculation of the dark matter annihilation cross sections}
%\label{DM}
%%%%%%%%%%%

In this model structure, the DM candidate interacts with the $Z^\prime$ boson to correctly manifest the bounds on the relic abundance. 
The coupling between the DM candidate and the $Z^\prime$ boson will be governed by the DM's U$(1)_X$ change $n_\chi$. 
On the other hand the $Z^\prime$ will be produced and decay into the SM particles and these vertices will be governed by the respective U$(1)_X$ charges of the SM particles. 
The DM annihilation processes are shown in Fig.~\ref{DM}. We calculate the cross sections of these processes in the center of mass (CM) frame in the following way : 
%%%%%%%%%%%%%
\begin{figure}[h]
\centering
\includegraphics[width=0.8\textwidth]{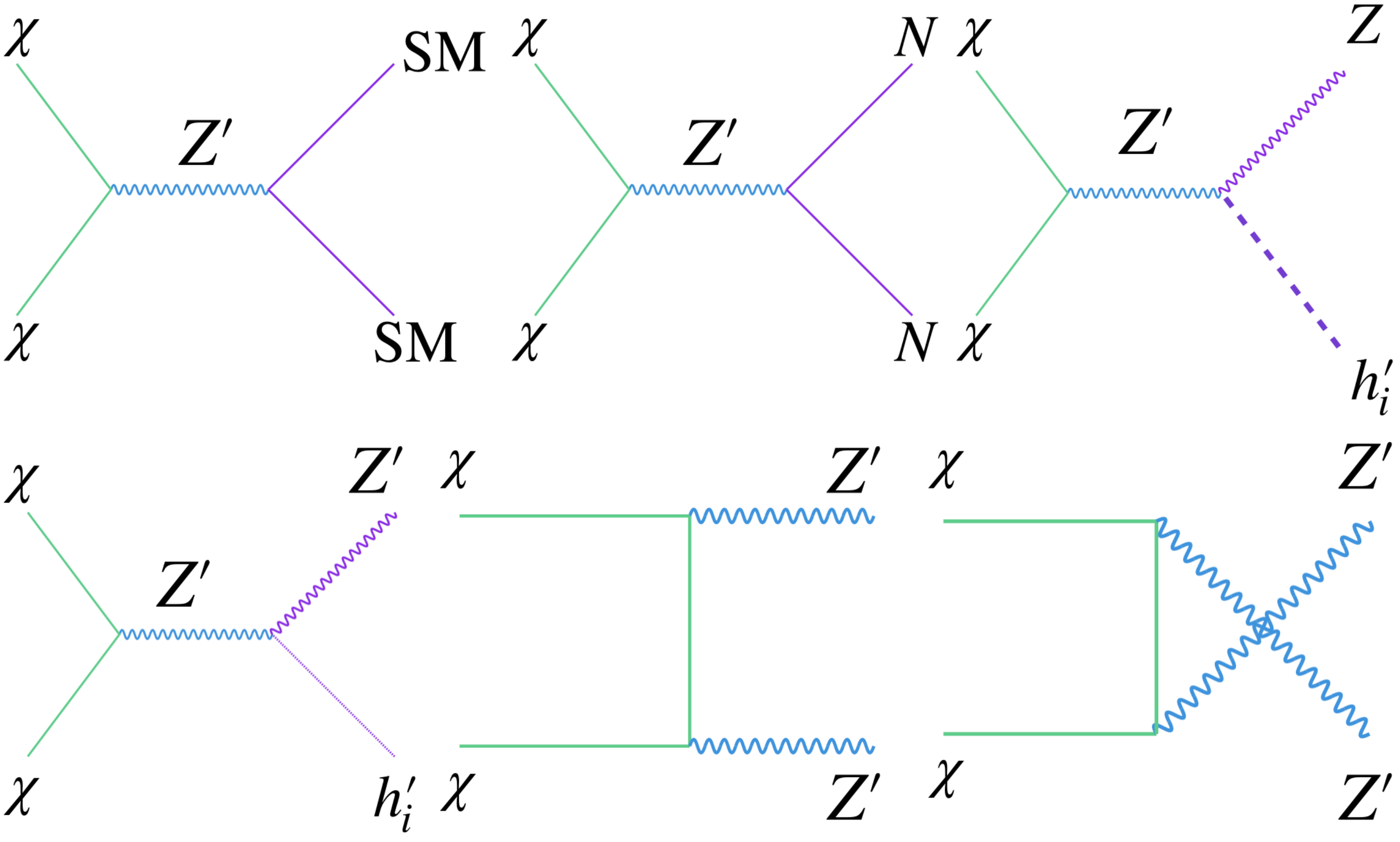} 
\caption{DM annihilation processes. SM denotes the SM fermions, $N$ denotes the right anded neutrinos and $h_i^\prime$ denotes for the two physical mass eigenstates of scalars whose mixing angle is $\beta$.}
\label{DM}
\end{figure}
%%%%%%%%%%%%%%%%%%%%%%%
%\subsection{$\chi \di{\chi} \to f \di{f}$ }
%%%%%%%%%%%%%%%%%%%%%
\begin{itemize}
\item[(i)] The interesting $s$- channel DM annihilation process into the SM fermions can be found from the $s$- channel process $\chi \di{\chi} \to f \di{f}$. 
The vertex between the fermions and the $Z^\prime$, namely $f \di{f} Z^\prime$ is given by
%\bal
$\mathcal{L}_{f\di{f}Z^\prime} = S_f g_X \di{f} \gamma^\mu ( n_V + n_A \gamma_5 ) f Z^\prime_{\mu}.$
%\eal
where $S_f$ is the symmetry factor and it is $1$ for the Dirac and $\frac{1}{2}$ for the Majorana fermions respectively.
Hence the Feynman rule for this vertex can be give by $i g_X \gamma^\mu (n_V + n_A \gamma_5)$ for both of the Dirac and Majorana fermions. 
The coefficients $n_V$ and $n_A$ for each fermion are shown in Table~\ref{table:cV_and_cA}.
Note that $n_V=0$ for $\nu_i$ and $N_i$, because their Majorana nature.
%%%%%%%%%%%%%%%%%%%%%%%
\begin{table}
\begin{center}
\small
\begin{tabular}{|c|c|c|c|c|c|} \hline 
 & $u_i$ & $d_i$ & $\ell_i$ & $\nu_i$ & $N_i$ \\ \hline \hline
$n_V$& $\frac{ 5 }{ 12 } x_H + \frac{ 1 }{ 3 } x_\Phi$ & $-\frac{ 1 }{ 12 } x_H + \frac{ 1 }{ 3 } x_\Phi$ & $-\frac{ 3 }{ 4 } x_H - x_\Phi$ & 0 & 0 \\ \hline \hline
$n_A$  & $\frac{ 1 }{ 4 } x_H$  & $ - \frac{ 1 }{ 4 } x_H $ & $- \frac{ 1 }{ 4 } x_H$ & $\frac{ 1 }{ 2 } x_H + x_\Phi$ & $ - x_\Phi$ \\ \hline \hline 
\end{tabular}
\caption{The coefficients $n_V$ and $n_A$ for each fermion.}
\label{table:cV_and_cA}
\end{center}
\end{table}
%%%%%%%%%%%%%%%%%%%%%
The scattering cross section in the CM frame is given by
\bal
\label{eq:sigmaff}
\sigma_{f\di{f}}
=  \frac{ S_f N_c }{ 12 \pi s }  \sqrt{ \frac{ s - 4 m_f^2 }{ s - 4 m_{DM}^2 } }
\frac{ n_\chi^2 g_X^4  ( s + 2 m_{DM}^2 ) }{ ( s - m_{Z^\prime}^2 )^2 + m_{Z^\prime}^2 \Gamma_{Z^\prime}^2 }
\Bigl\{ n_V^2 ( s + 2 m_f^2 ) + n_A^2 ( s - 4 m_f^2 ) \Bigr\}.
\eal
%%%%%%%%%%%%%%%%%%%
%\subsection{$\chi \di{\chi} \to Z h_i^\prime\: (i = 1, 2)$} 
%%%%%%%%%%%%%%%%%%%

\item[(ii)] The vertex between the scalars $(h_i^\prime, i=1, 2)$, $Z$ and $Z^\prime$ is given by
%\bal
%\label{eq:sigmahZ}
$\mathcal{L}_{h_i^\prime Z Z^\prime}
= - (m_Z x_H g_X \beta_i) h_i^\prime Z^\mu Z^\prime_\mu$, 
%\eal
where $(\beta_1, \beta_2) = ( \cos\beta, \sin\beta )$ and $m_Z$ is the SM $Z$ boson mass. 
The scattering cross section of this $s$- channel annihilation process $\chi \di{\chi} \to Z h_i^\prime\: (i = 1, 2)$  in the CM frame can be given by
\bal
\sigma_{h_i^\prime Z}
= &\frac{ n_\chi^2 x_H^2 \beta_i^2 g_X^4 }{ 196 \pi s} 
\left( 1 + \frac{ 2 m_{DM}^2 }{ s } \right) 
\frac{ s^2 + 2 s ( 5 m_Z^2 - m_{h_i^\prime}^2 ) + (m_Z^2 - m_{h_i^\prime}^2)^2}
{(s-m_{Z^\prime} )^2 + m_{Z^\prime}^2 \Gamma_{Z^\prime}^2 }
\nonumber \\
&\times \sqrt{ \frac{s^2 -2s(m_Z^2 + m_{h_i^\prime}^2) + (m_Z^2 - m_{h_i^\prime}^2)^2}{s ( s - 4 m_{DM}^2 )} }.
\eal
%%%%%%%%%%%%%%%%%%%%%%%%%%%%%%%%%%%
%\subsection{$\chi \di{\chi} \to Z^\prime h_i^\prime\: (i=1, 2)$ }
%%%%%%%%%%%%%%%%%%%%%%%%%%%%%%%%%%%%

\item[(iii)] The vertex between the scalars $(h_i^\prime, i = 1, 2)$ and  $Z^\prime$ is given by
%\bal
$\mathcal{L}_{h_i^\prime Z^\prime Z^\prime} = \frac{ 1 }{ 2 } g_X \sigma_i Z^{\prime\mu} Z^\prime_\mu h_i^\prime$,
%\eal
where $\sigma_i\: (i=1, 2)$ is dimension-one coupling constant;
%\bal
$\sigma_1 = g_X ( \frac{ x_H^2 }{ 2 } v_H \cos \beta+$ $8 x_\Phi^2 v_\Phi \sin \beta )$ 
and $\sigma_2 = g_X ( - \frac{ x_H^2 }{ 2 } v_H \sin \beta + 8 x_\Phi^2 v_\Phi \cos \beta )$. 
%\eal
Hence the $s$- channel scattering cross section for the process $\chi \di{\chi} \to Z^\prime h_i^\prime\: (i=1, 2)$ in the CM frame is given by
\bal
\label{eq:sigmahZ'}
\sigma_{h_i^\prime Z^\prime}
= & \frac{ \xi_\beta n_\chi^2 g_X^4 }{ 48 \pi s } \left( 1 + \frac{ 2 m_{DM}^2 }{ s } \right)
\frac{ s^2 + 2 s ( 5 m_{Z^\prime}^2 - m_{h_i^\prime}^2 ) + ( m_{Z^\prime}^2  - m^2_{h_i^\prime} )^2 }{ ( s - m_{Z^\prime}^2 )^2 + m_{Z^\prime}^2 \Gamma_{Z^\prime}^2 }
\nonumber \\
& \times \sqrt{ 
		\frac{ s^2 - 2 s ( m_{Z^\prime}^2 + m_{h_i^\prime}^2 ) 
			+ (  m_{Z^\prime}^2 - m_{h_i^\prime}^2)^2 }
			{ s ( s - 4 m_{DM}^2 )}
		},
\eal
where $\xi_i (i=1, 2)$ is defined as $\xi_i = \sigma_i^2 / (4 m_{Z^\prime}^2 )$ and can be written as $\xi_1 \sim 4 \sin^2 \beta + \frac{ v_H }{ 4 v_\Phi} x_H^2 \sin(2\beta)$ and $\xi_2 \sim 4 \cos^2 \beta - \frac{ v }{ 4 v_\Phi} x_H^2 \sin(2\beta)$ using $v_\Phi \gg v_H$.
%%%%%%%%%%%%%%%%%%%%%%%%%%%%%
%\subsection{$ \chi \di{ \chi } \to Z^\prime Z^\prime$ }
%%%%%%%%%%%%%%%%%%%%%%%%%%%%%%%%

\item[(iv)] The $ \chi \di{ \chi } \to Z^\prime Z^\prime$ process includes the $t$- channel and $u$- channel processes and their interference. 
The scattering cross section is give by in the CM frame is given by 
\bal
\label{eq:sigmaZ'Z'}
\sigma_{Z^\prime Z^\prime}
= &
\nonumber \\
& \frac{ n^4 g_X^4 }{ 8 \pi s } \frac{ \omega }{ v }
\Biggl\{
	-1 - \frac{ ( 2 + r^\prime_z{}^2 )^2 }{ (2-r^\prime_z{}^2)^2 + 4v^2 } 
	+ \frac{ 6 + 12 v^2 + 4v^4 -2r^\prime_z{}^2 +r^\prime_z{}^4 }{ 2v\omega ( 1 + v^2 + \omega^2 )}
	\ln\left( \frac{ 1 + ( v + \omega )^2 }{ 1 + ( v - \omega )^2 } \right)
\Biggr\},
\eal
where $r_z$, $\omega$ and $v$ can be defined as 
%\bal
$r^\prime_z = \frac{m_{Z^\prime} }{ m_{DM}^{}}, \: v = \frac{p}{m_{DM}^{}}, \: 
\omega = \frac{ k }{ m_{DM}^{}}$.
%\eal
There is a typographical error in the denominator of the first term of Eq.~(\ref{eq:sigmaZ'Z'}) with Ref.~\cite{FileviezPerez:2018toq}.
Using Eqn.~\ref{eq:sigmaff}-\ref{eq:sigmaZ'Z'} we calculate the thermal averaged cross sections to estimate the correct DM relic abundance. 
\end{itemize}
%%%%%%%%%%%%%%%%%%%%%%%%%%%%%%%%%%
%\section{Dark matter phenomenology}
%\label{calc}
%%%%%%%%%%%%%%%%%%%%%%%%%%%%%%%

The DM relic abundance $\Omega h^2$ has been measured as $0.120\pm0.001$ by the Planck satellite \cite{Aghanim:2018eyx}. 
In the context of our model we calculate the thermal averaged cross sections using \cite{Gondolo:1990dk} for the different DM annihilation processes
\bal
\label{eq:1}
& \bigl< \sigma_{f \di{f}} v_M \bigr> = 
	 \frac{ S_f N_C n_\chi^2 g_X^4 }{ 2\pi m_{DM}^2 }
	\frac{ \sqrt{ 1 - r_f^2 } }{ ( 4 - r_z^\prime{}^2)^2 + r_z^\prime{}^4\Gamma_{Z^\prime}^2/m_Z^2 }
	\Bigl\{ n_V^2 (2 + r_f^2 ) + 2 n_A^2 ( 1 - r_f^2 ) \Bigr\},
\\[15pt]
\label{eq:2}
& \bigl< \sigma_{h_i^\prime Z} v_M \bigr> = 
	\frac{  n_\chi^2 x_H^2 \beta_i^2 g_X^4 }{ 1024 \pi m_{DM}^2 }
	\frac{ \sqrt{ (r_z^2-r_{h_i^\prime}^2)^2+ 8(2 - r_z^2 -r_{h_i^\prime}^2)^2} }
		{ ( 4 - r_z^\prime{}^2 ) + r_z^\prime{}^4 \Gamma_{Z^\prime}^2/m_{Z^\prime}^2 }
	\Bigl\{ 
		16 + 8 ( 5 r_z^2 - r_{h_i^\prime}^2 ) + ( r_z^2 - r_{h_i^\prime}^2 )^2
	\Bigr\},
\\
\label{eq:3}
& \bigl< \sigma_{h_i^\prime Z^\prime} v_M \bigr> = 
	\frac{ \xi_i n_\chi^2 g_X^4 }{ 256 \pi m_{DM}^2 }
	\frac{ \sqrt{ (r_z^\prime{}^2-r_{h_i^\prime}^2)^2+ 8(2 - r_z^\prime{}^2 -r_{h_i^\prime}^2)^2 } }
		{ ( 4 - r_z^\prime{}^2 ) + r_z^\prime{}^4 \Gamma_{Z^\prime}^2/m_{Z^\prime}^2 }
	\Bigl\{ 
		16 + 8 ( 5 r_z^\prime{}^2 - r_{h_i^\prime}^2 ) + ( r_z^\prime{}^2 - r_{h_i^\prime}^2 )^2
	\Bigr\},
\\
\label{eq:4}
& \bigl< \sigma_{Z^\prime Z^\prime} v_M \bigr> = 
	\frac{ n_\chi^4 g_X^4 }{ 4 \pi m_{DM}^2 }
	\frac{ ( 1 - r_z^\prime{}^2 )^{\frac{3}{2}} }{ ( 2 - r_z^\prime{}^2 )^2 },
\eal
where $r_f = \frac{ m_f }{ m_{DM}^{} }, r_{h_i^\prime} = \frac{ m_{h_i^\prime} }{ m_{DM}^{} }, r_z = \frac{ m_Z }{ m_{DM}^{} }$ and $r_z^\prime = \frac{ m_{Z^\prime} }{ m_{DM}^{} }$. In this analysis $v_M$ is considered to be the M$\slashed{o}$ller velocity which can be evaluated as $v_M=2\sqrt{ \frac{ s - 4 m_{DM}^2 }{ s } }$ in the CM frame. Therefore the total thermal averaged cross section can be written as 
\bea
\bigl< \sigma v_M \bigr> =\sum_{f} \bigl< \sigma_{f \di{f}} v_M \bigr> + \sum_{i=1}^2 \bigl< \sigma_{h_i^\prime Z^\prime} v_M \bigr> +  \bigl< \sigma_{Z^\prime Z^\prime} v_M \bigr>, 
\label{eq:5}
\eea
which is a function of $x_H$ and $n_\chi$. We define a ratio $\frac{\bigl<\sigma v_M\bigr>_i}{\bigl<\sigma v_M\bigr>}$ which dependents on $x_H$ and $n_\chi$. The numerator $\bigl<\sigma v_M\bigr>_i$ comes from each mode of the DM annihilation written in Eqs.~(\ref{eq:1})-(\ref{eq:4}) whereas the denominator is the sum of all the modes as written in Eq.~(\ref{eq:5}). 

\begin{figure}[h]
\centering
\includegraphics[width=0.325\textwidth]{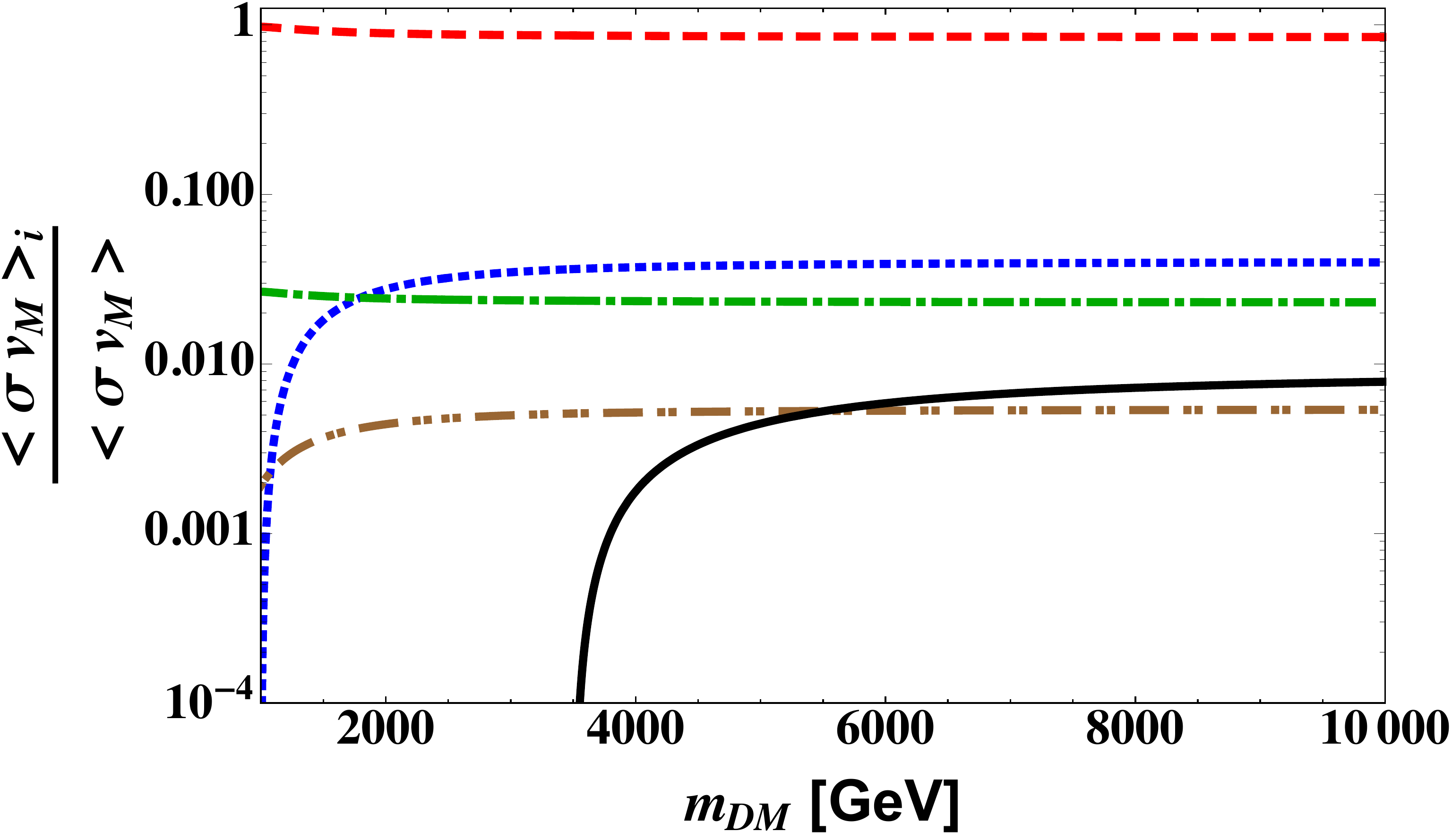} 
\includegraphics[width=0.325\textwidth]{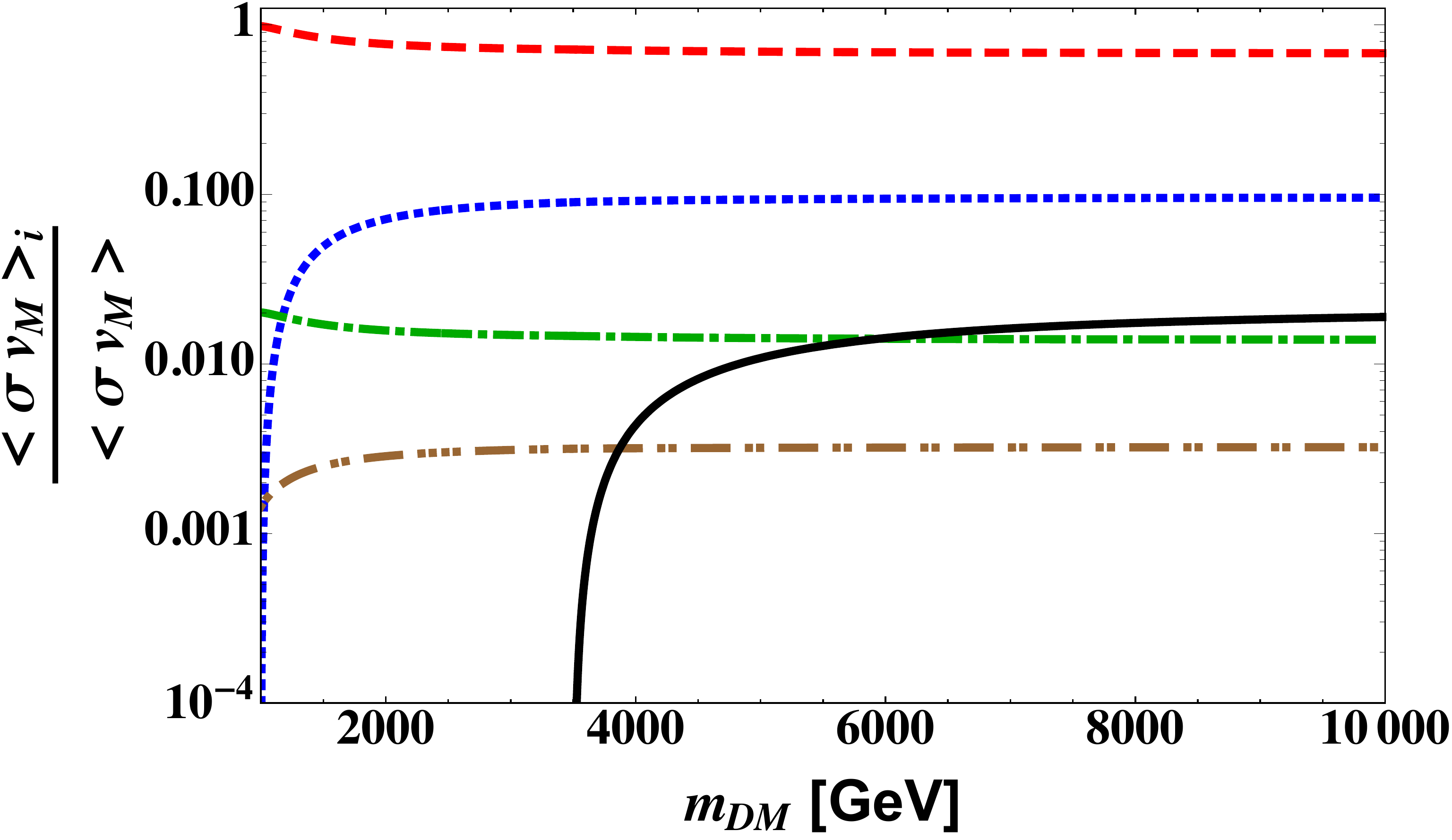} 
\includegraphics[width=0.325\textwidth]{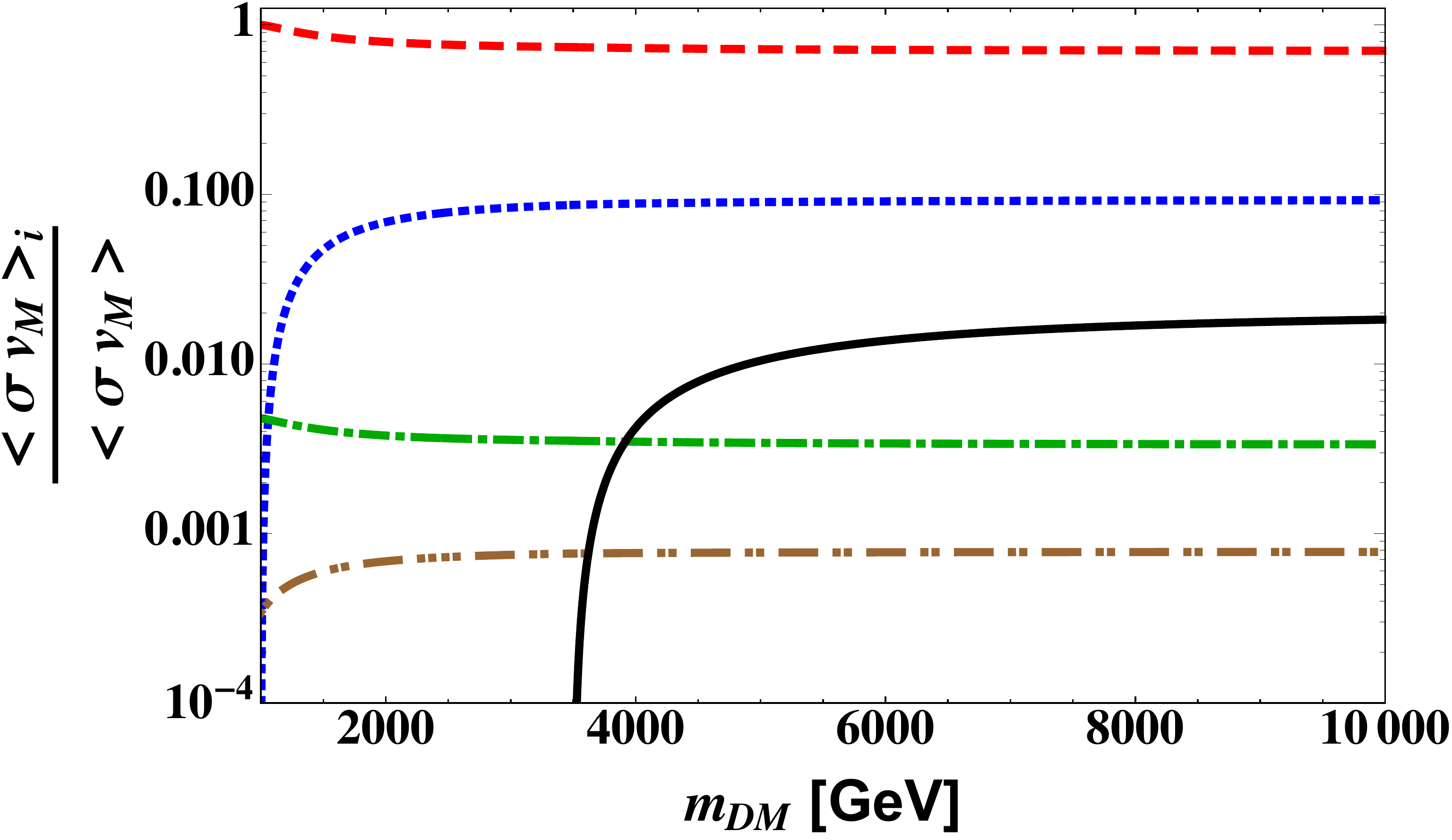}\\
\includegraphics[width=0.325\textwidth]{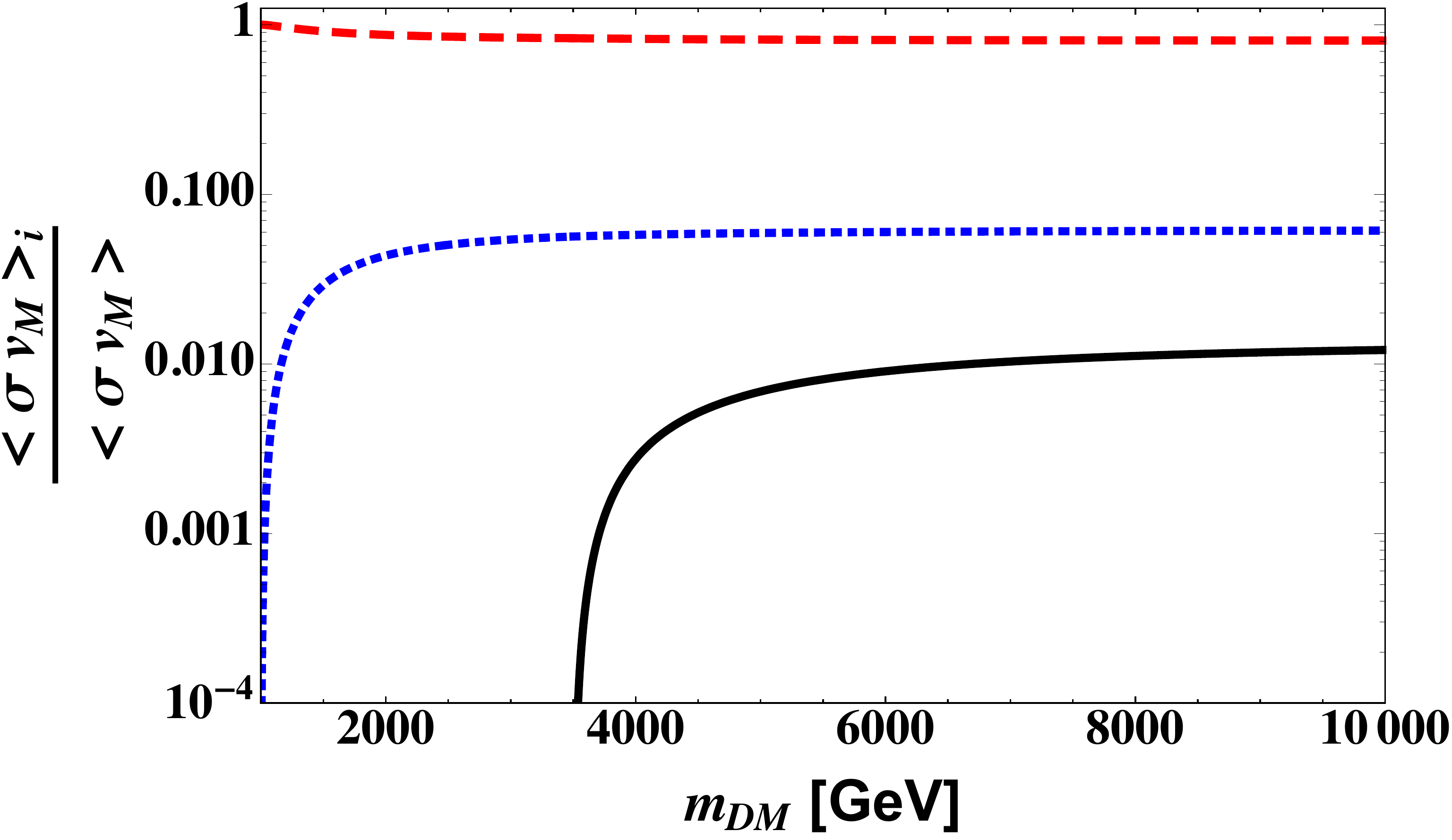} 
\includegraphics[width=0.325\textwidth]{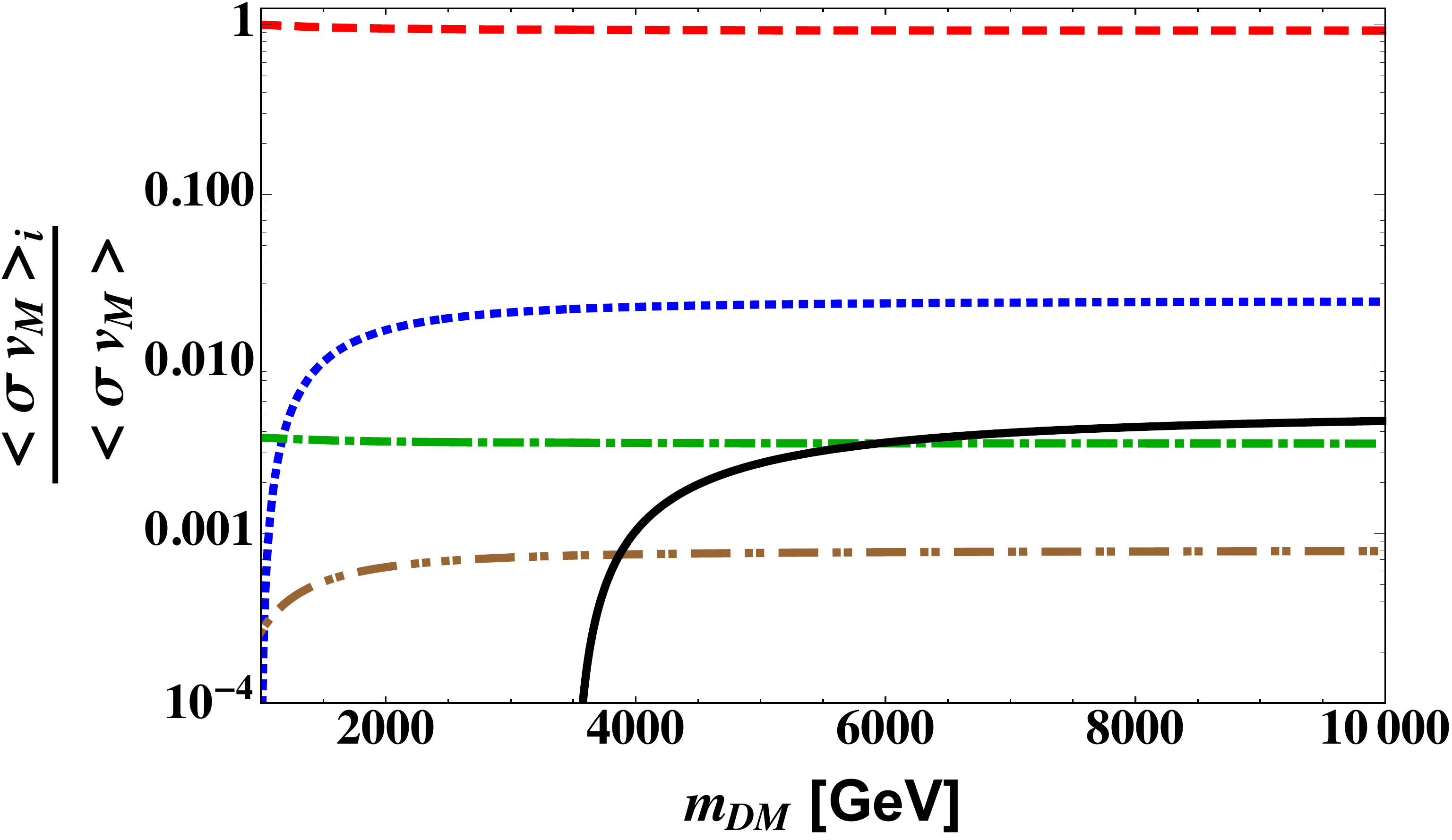} 
\includegraphics[width=0.325\textwidth]{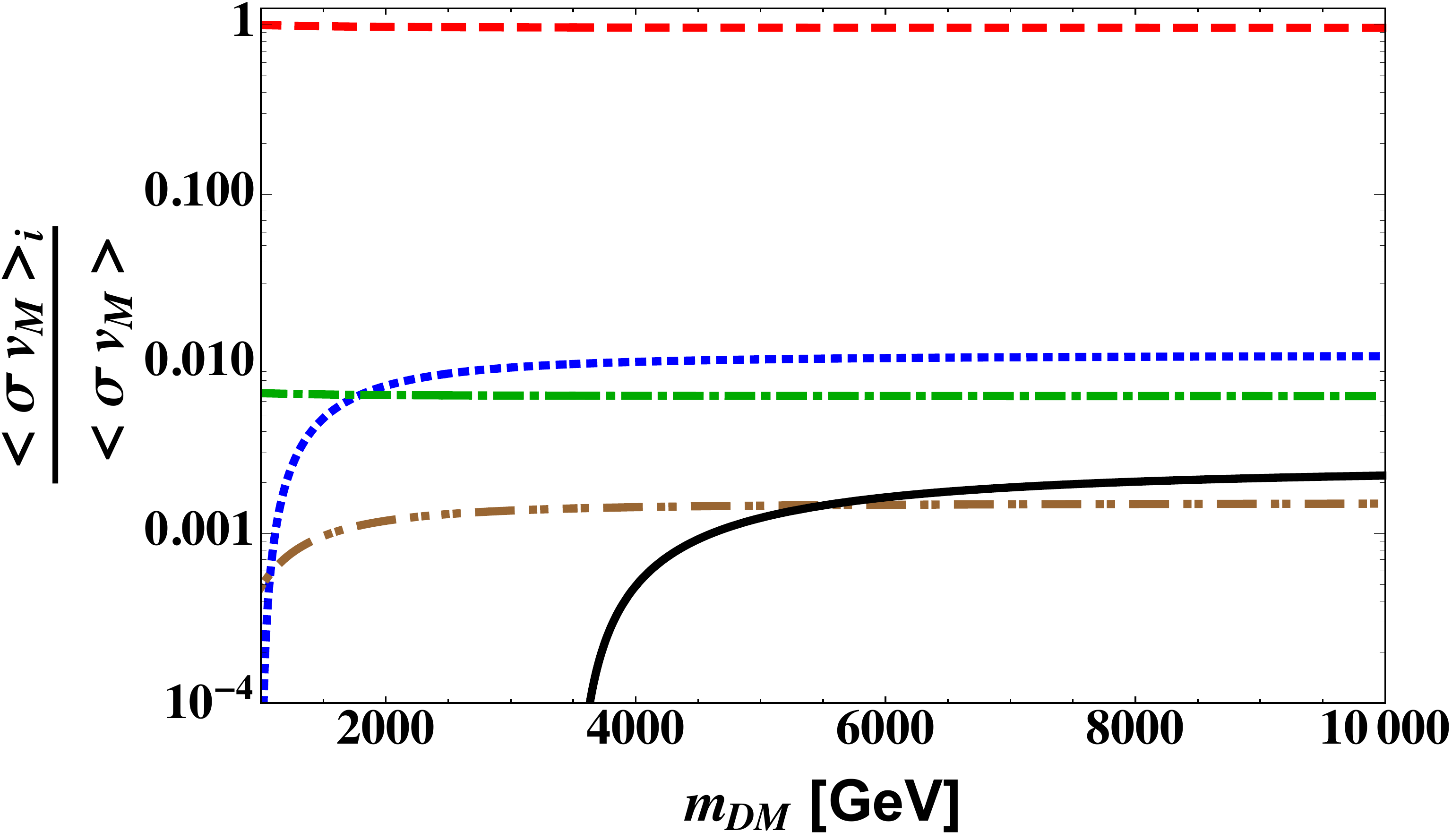}
\caption{The ratio $\frac{\bigl<\sigma v_M\bigr>_i}{\bigl<\sigma v_M\bigr>}$ as a function of $m_{\rm{DM}}$ for $n_\chi=\frac{1}{3}$ with $x_H= -2, -1, -0.5$ in the top panel from left to right and the same with $x_H= 0, 1, 2$ in the bottom panel from left to right. We present the SM lepton pair by the red dashed line, right handed neutrino pair by blue dashed line, $Z h_1^\prime$ by the darker green dot-dashed line, $Z h_2^\prime$ by the brown double-dot dashed line and $Z^\prime Z^\prime$ by black solid line. In this analysis we consider $m_{Z^\prime}=3.5$ TeV, $m_{h_1^\prime}=125$ GeV and $m_{h_2^\prime}=600$ GeV, respectively.}
\label{sigV1}
\end{figure}

\begin{figure}[h]
\centering
\includegraphics[width=0.325\textwidth]{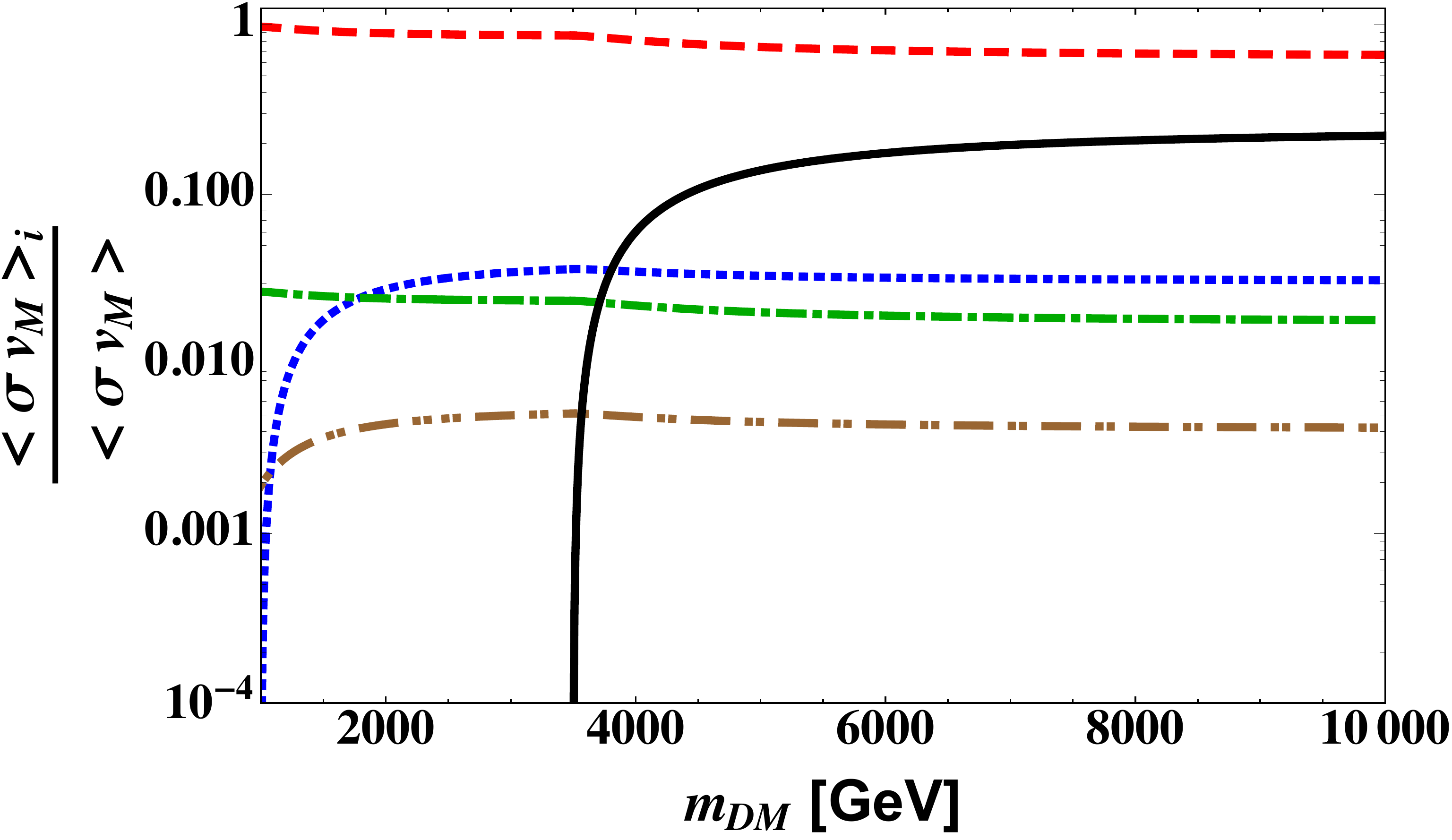} 
\includegraphics[width=0.325\textwidth]{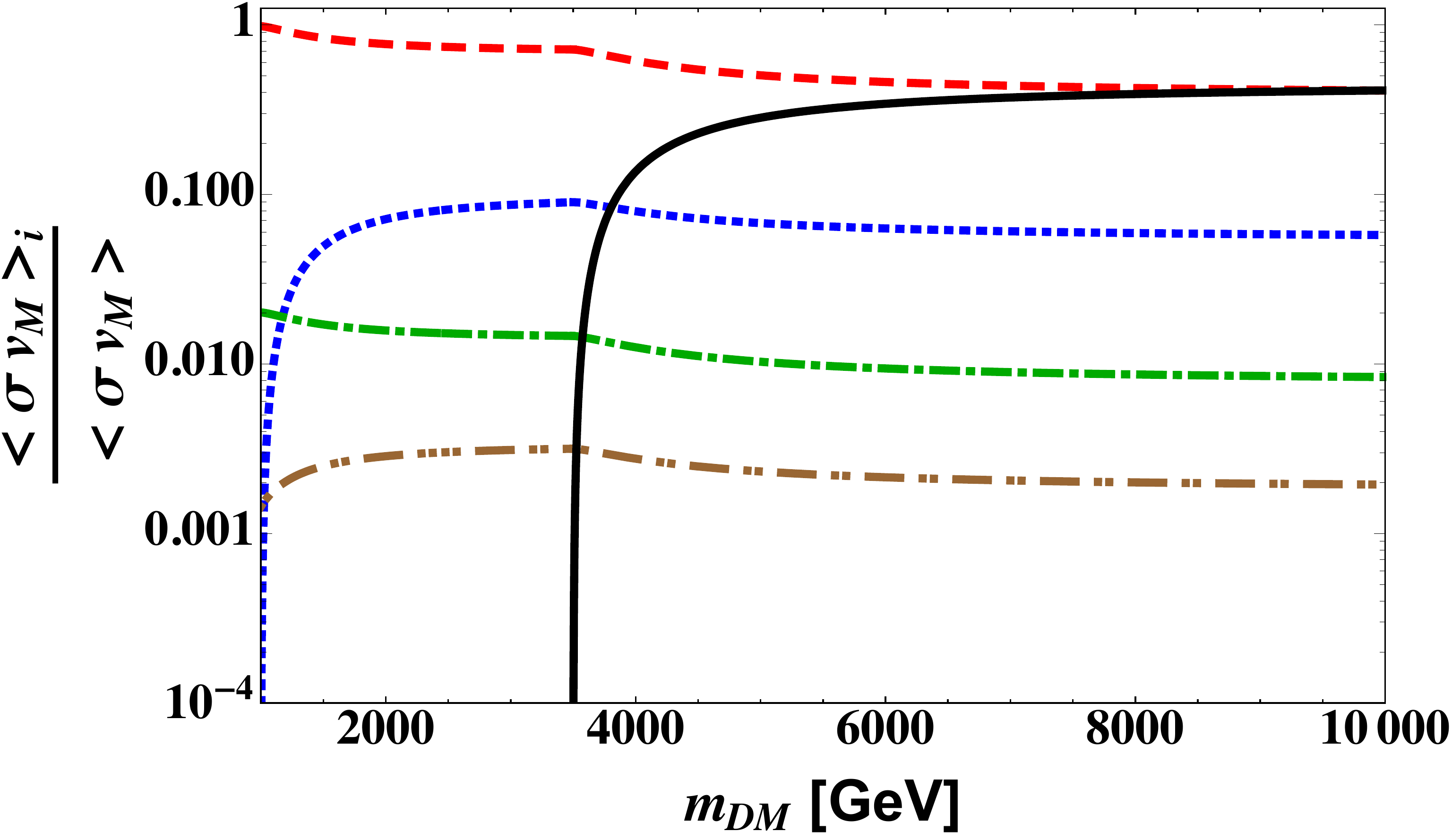} 
\includegraphics[width=0.325\textwidth]{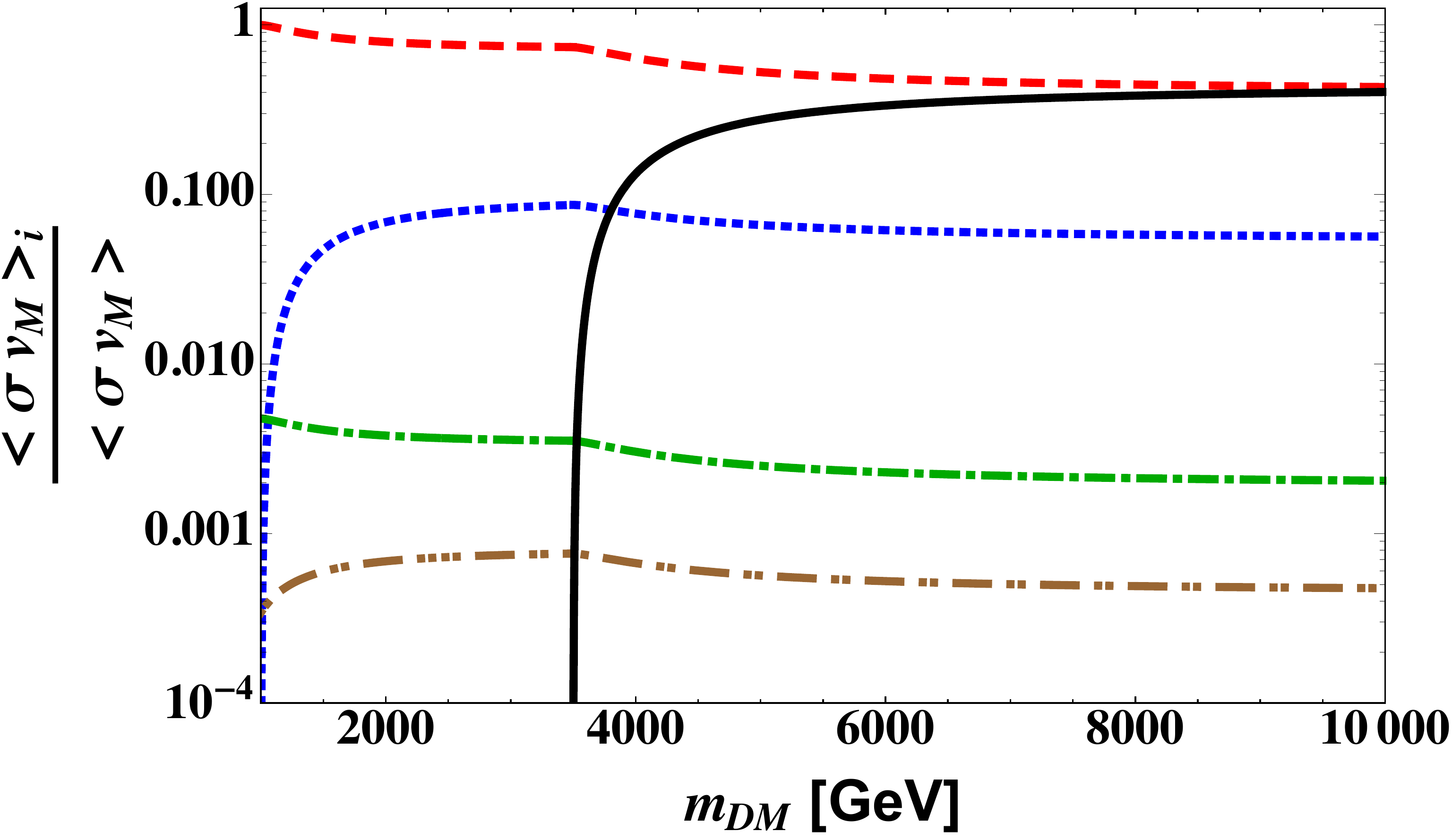}\\
\includegraphics[width=0.325\textwidth]{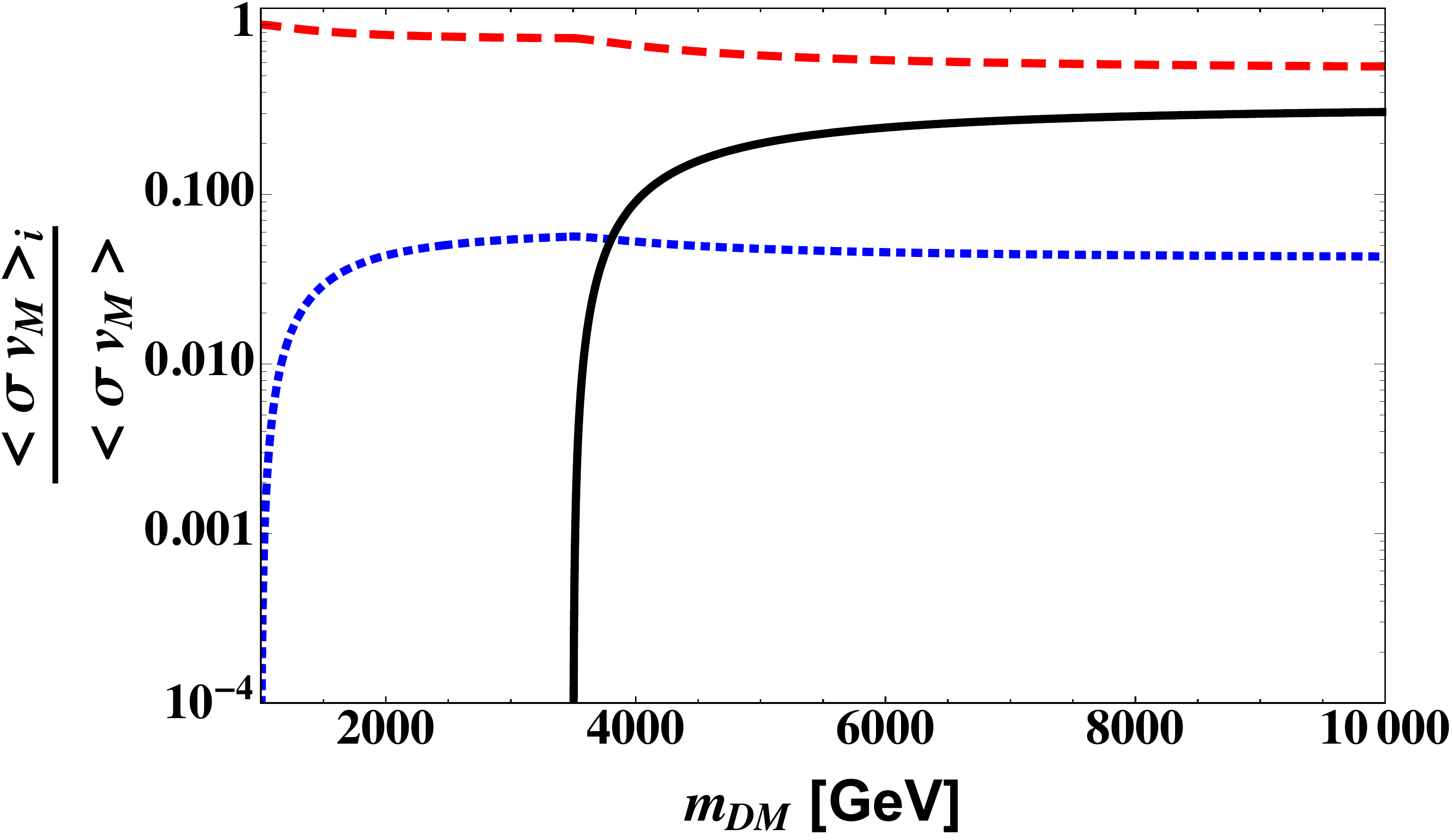} 
\includegraphics[width=0.325\textwidth]{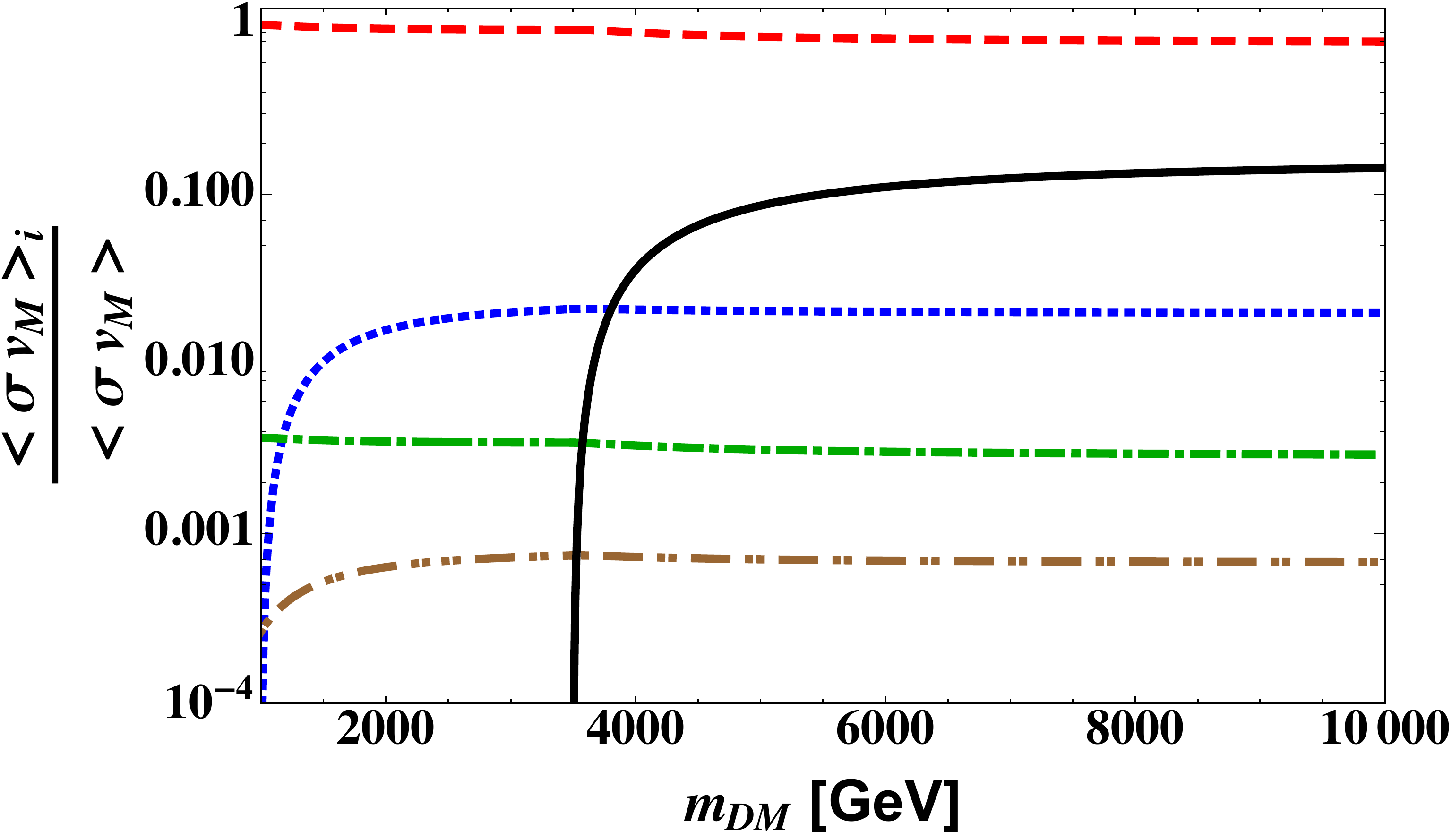} 
\includegraphics[width=0.325\textwidth]{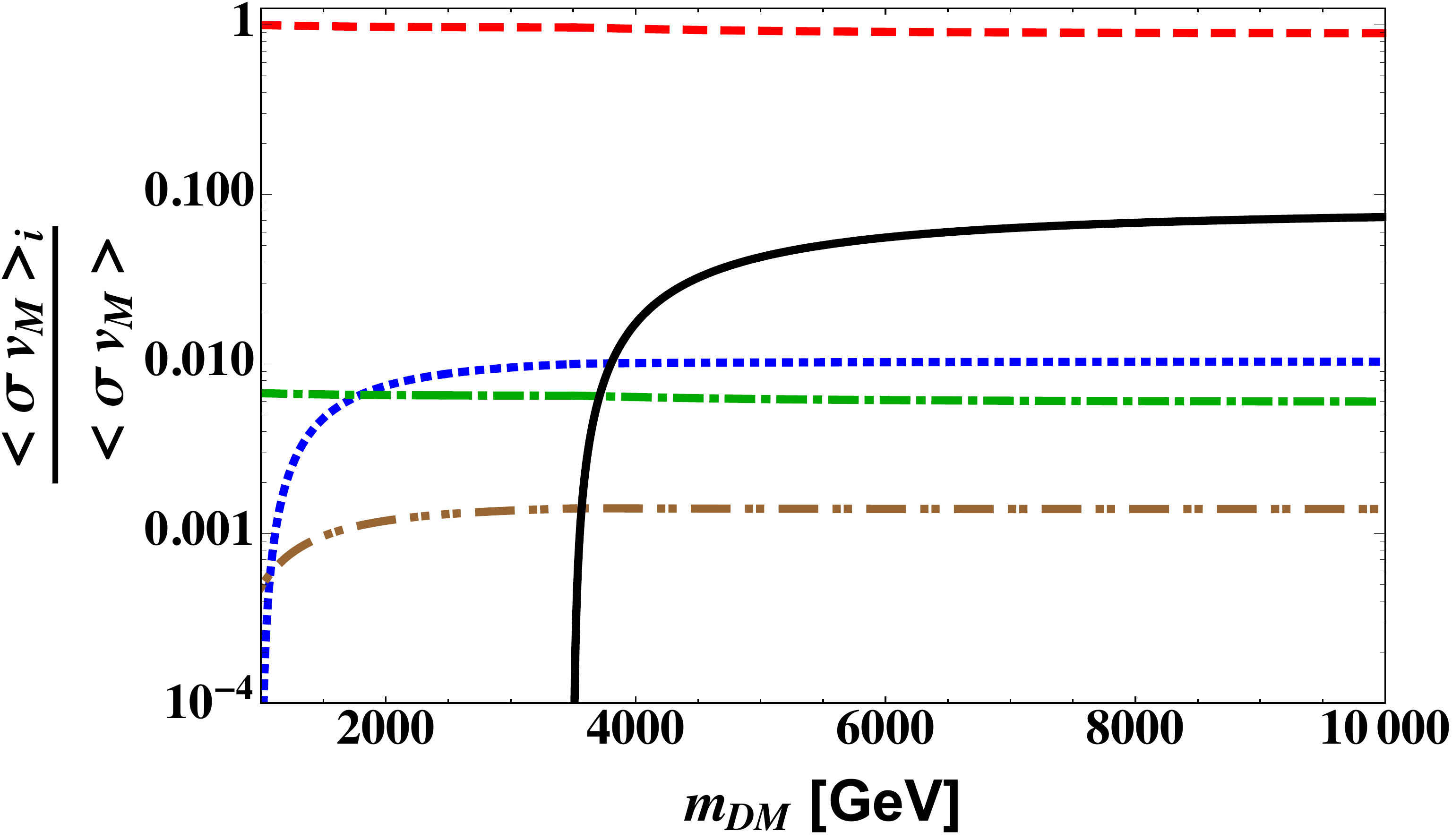}
\caption{Same as Fig.~\ref{sigV1} but $n_\chi=2$.}
\label{sigV2}
\end{figure}

We plot this ratio with respect to the DM mass $(m_{\rm{DM}})$ in Figs.~\ref{sigV1} and \ref{sigV2} for two different values of $n_\chi$ such as $n_\chi=\frac{1}{3}$ and $n_\chi=2$  for different choices of $x_H= -2, -1, -0.5, 0, 1, 2$, respectively. In this analysis we have used the Taylor expansion of the Mandelstum variable $s$ and took the leading terms in the calculation of the thermally averaged cross section \cite{Gondolo:1990dk}. For simplicity we have calculated the cross sections in the CM frame.
%%%%%%%%%%%%%%%%%%%%%%%%%%%%%%%%%%%%%%%%
%%%%%%%%%%%%%%%%%%%
%%%%%%%%%%%%%%%%%%%
In these figures we represent the SM lepton pair by the red dashed line, right handed neutrino pair by blue dashed line,
$Z h_1^\prime$ by the darker green dot-dashed line, $Z h_2^\prime$ by the brown double-dot dashed line and $Z^\prime Z^\prime$ by black solid line. 
In the top panel of the Figs.~\ref{sigV1} and \ref{sigV2} from left to right we have $x_H= -2, -1, -0.5,$ respectively and in the bottom panel of the same figure from left to right we have $x_H= 0, 1, 2,$ respectively. In this analysis we consider $m_{Z^\prime}=3.5$ TeV and $g^\prime=0.1$ which satisfy the constraints obtained from the LHC as shown in. Fig.~\ref{bounds1}. We also consider $h_1^\prime$ as the SM Higgs fixing its mass as $m_{h_1^\prime}=125$ GeV where as the second Higgs mass is a free parameter, we fix its mass at $m_{h_2^\prime}=600$ GeV. We notice that the dominant contribution on the DM annihilation channel comes from the SM fermions $(f\overline{f})$. However, for the larger values of $n_\chi$ the $Z^\prime Z^\prime$ channel becomes comparable with the $f\overline{f}$ channel. We also find that the DM annihilation cross sections into $f\overline{f}$, $Zh^\prime_i$ and $Z^\prime h^\prime_i$ depend on $x_H$. We notice that the difference between the cases $x_H=-1$ and $-0.5$ is negligible and similar thing happens for $x_H=1$ and $2$. Therefore for the further study we concentrate on $x_H=-2, -0.5, 0$ and $2$.

The $s$- wave dominance is very prominent in these cases. Therefore the decoupling parameter $(x_f)$ \cite{Kolb:1990vq} can be evaluated by using the following formula 
\bea
x_f = \ln\Big[ 0.038 \frac{g}{\sqrt{g_\ast}} M_{\rm{Pl}} m_{\rm{DM}} \bigl<\sigma v_M\bigr>\Big]-\frac{1}{2} \ln\Big[\ln\Big[ 0.038 \frac{g}{\sqrt{g_\ast}} M_{\rm{Pl}} m_{\rm{DM}} \bigl<\sigma v_M\bigr>\Big]\Big]
\eea
where $g$ is the degrees of freedom of the DM candidate. Using the total thermally averaged cross section and $x_f$ finally we evaluate the DM relic abundance from \cite{Gondolo:1990dk} as
\bea
\Omega h^2 = 2\times1.04 \times 10^9 \frac{\sqrt{g_\ast}}{g_{\ast s}} \frac{x_f}{M_{\rm{Pl}} \bigl<\sigma v_M\bigr>} \rm{GeV}^{-1}
\label{relic}
\eea
where $g_\ast= g_{\ast s}=106.75$ as $\chi$ decouples at the radiation-dominated era. The factor of $2$ in from of Eq.~(\ref{relic}) comes from the $\chi$ and $\bar{\chi}$. 
Hence we identify the viable parameter regions of our model satisfying the upper bound on the relic abundance as $\Omega h^2 \leq 0.12$.

\begin{figure}[h]
\centering
\includegraphics[width=0.8\textwidth]{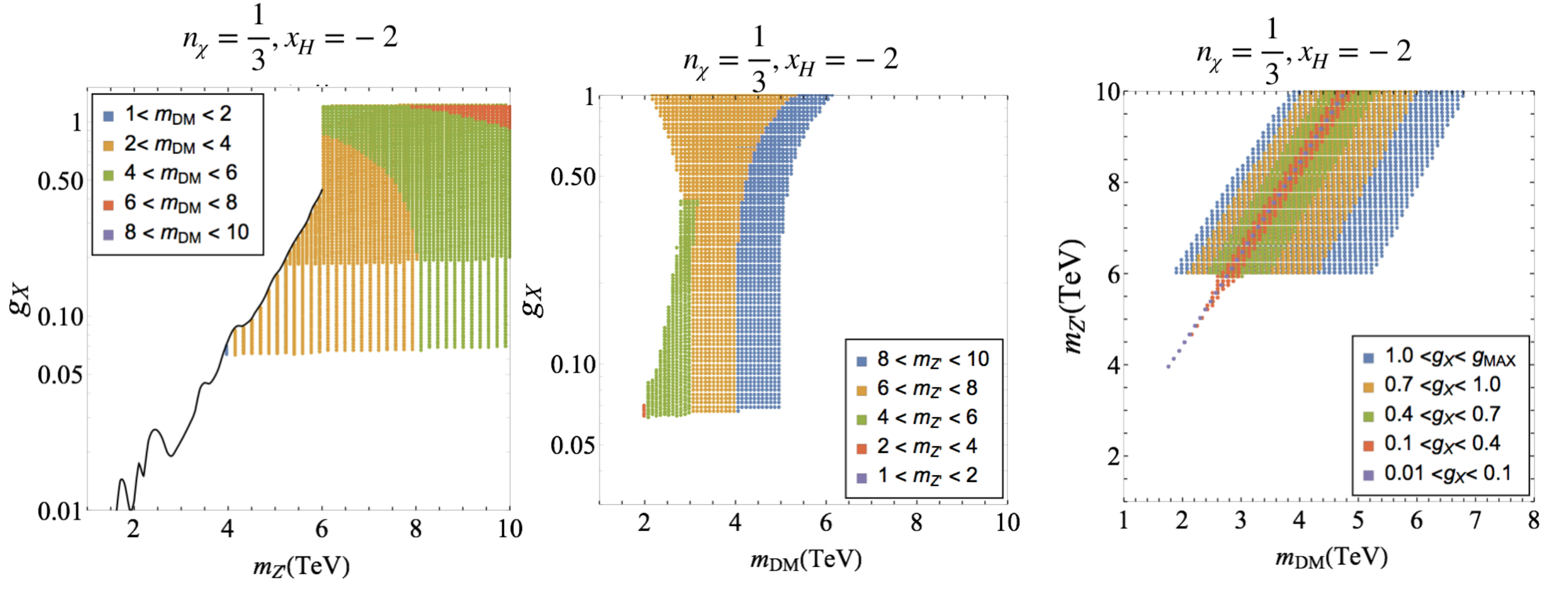} 
\includegraphics[width=0.8\textwidth]{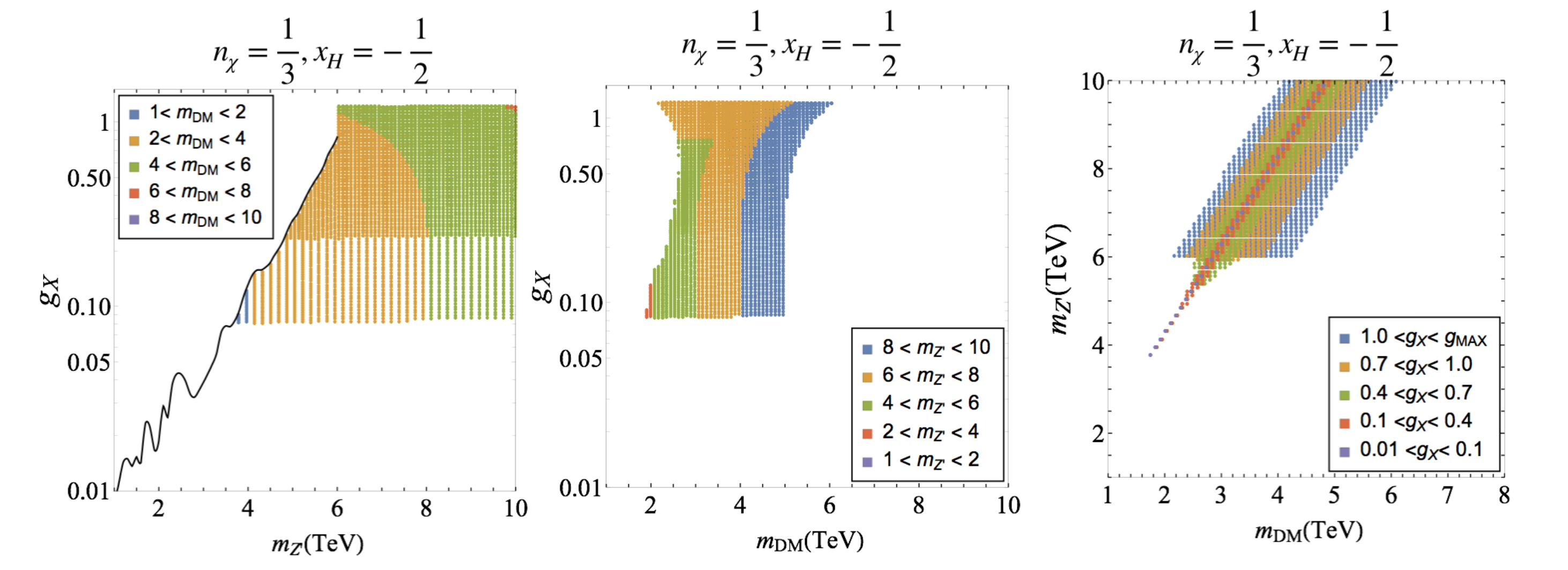} 
\caption{The constraints on the model parameters for $n_\chi=\frac{1}{3}$ and $x_H=-2$ (upper panel) and $n_\chi=\frac{1}{3}$ and $x_H=-\frac{1}{2}$ (lower panel).
The left panel shows the constraints on $g_X$ and $m_{Z^\prime}$ from the colliders (black, solid) and the constraints from the bound of the DM relic abundance.
The second panel shows the corresponding limits on $g_X$ and $m_{\rm{DM}}$ from the DM relic abundance satisfying the collider constraints. 
The third panel implies the limits on $m_{Z^\prime}$ and $m_{\rm{DM}}$ under the DM and collider constraints. 
The colors correspond to the different ranges of the model parameters.
}
\label{DM1-1}
\end{figure}

\begin{figure}[h]
\centering
\includegraphics[width=0.8\textwidth]{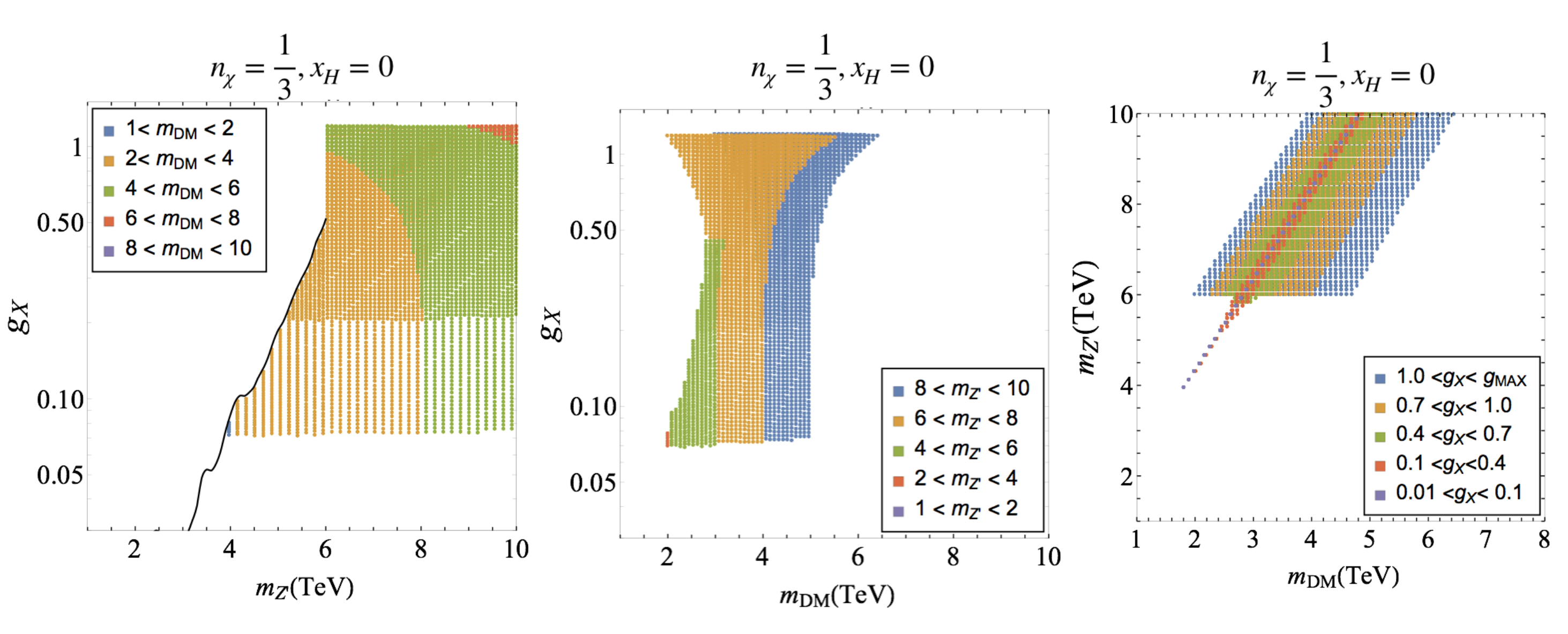} 
\includegraphics[width=0.8\textwidth]{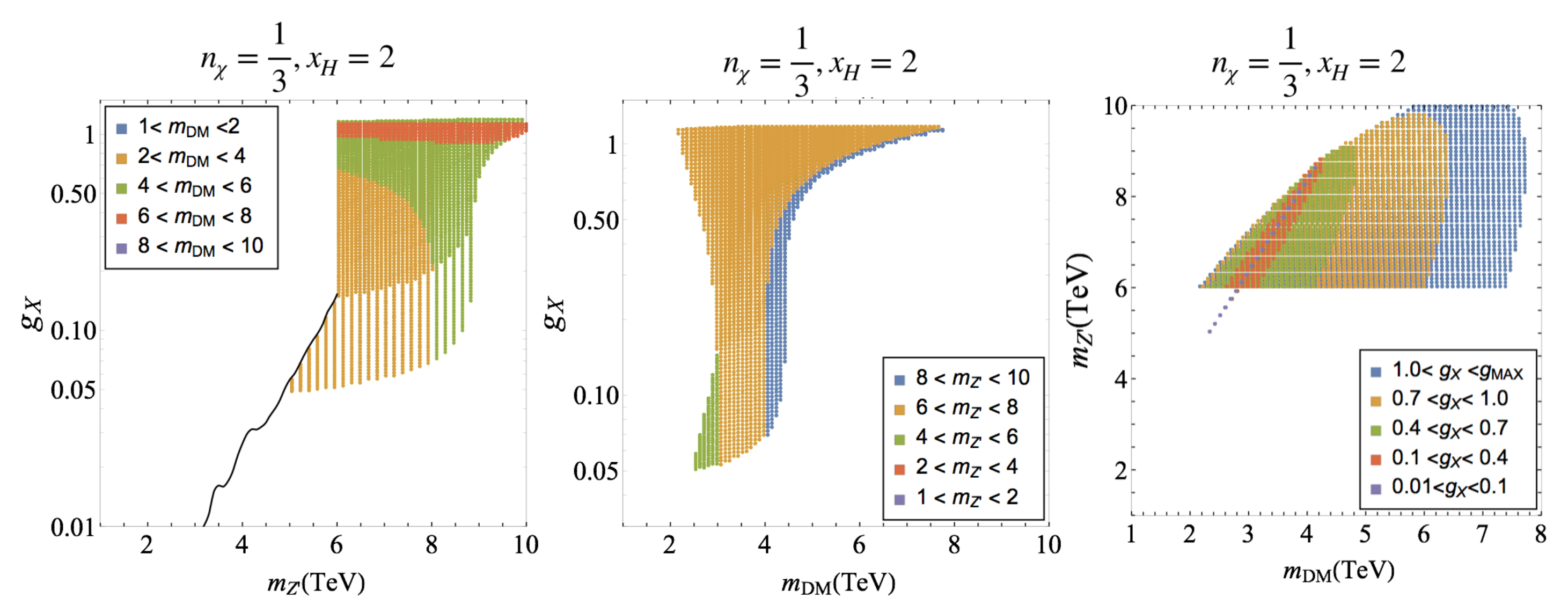} 
\caption{Same as Fig.~\ref{DM1-1} for $n=\frac{1}{3}$ with $x_H=0$ (upper panel) and $x_H=2$ (lower panel).}
\label{DM1-2}
\end{figure}

\begin{figure}[h]
\centering
\includegraphics[width=0.8\textwidth]{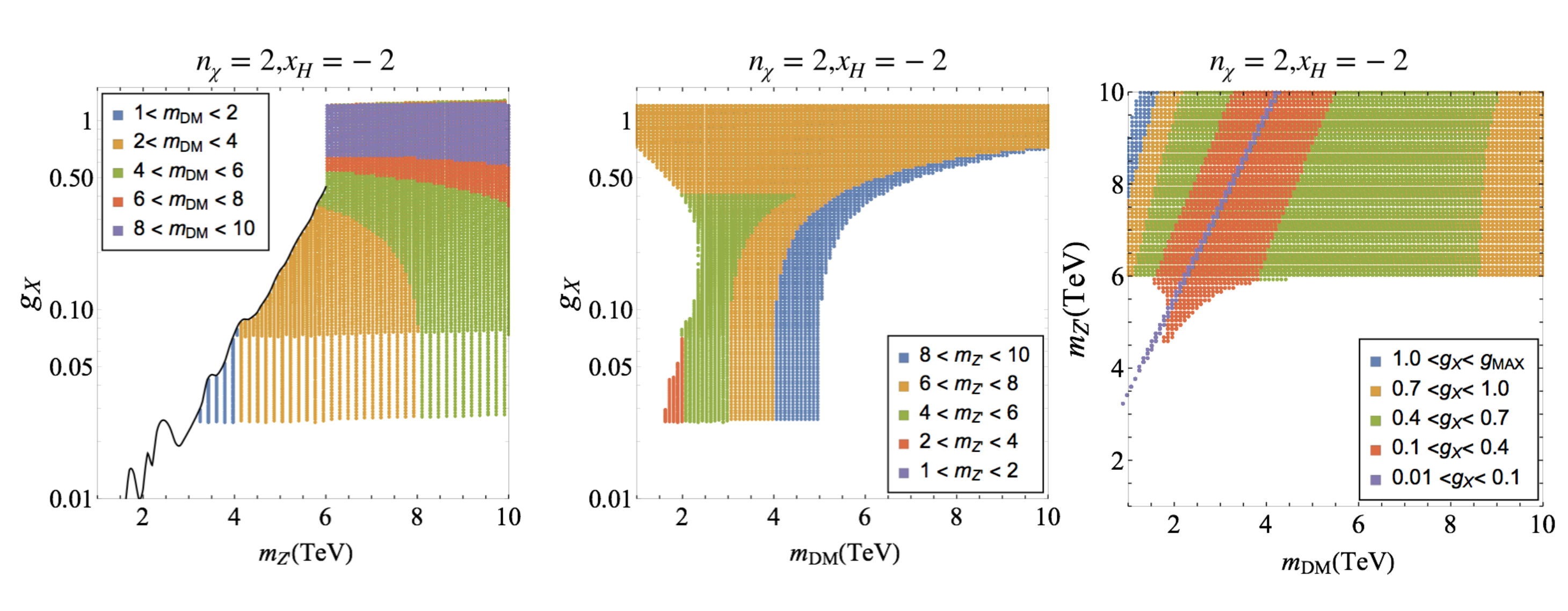} 
\includegraphics[width=0.8\textwidth]{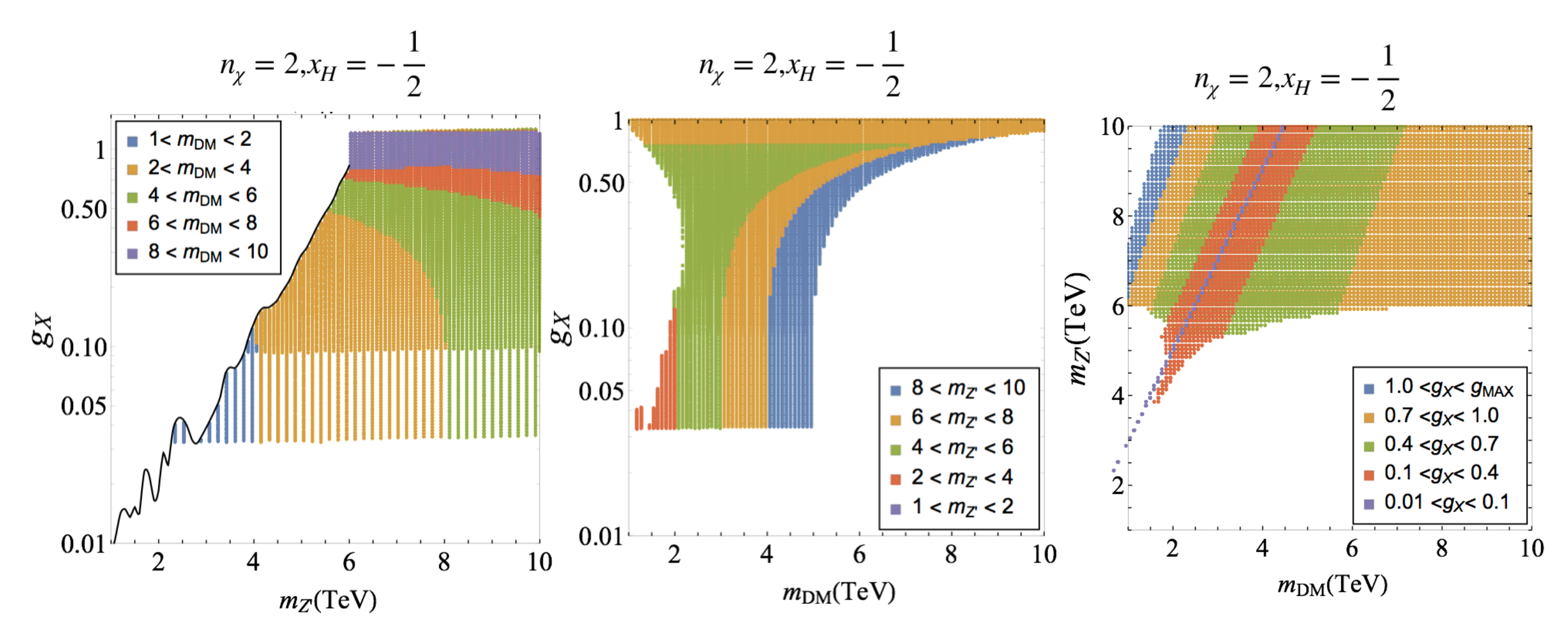} 
\caption{Same as Fig.~\ref{DM1-1} for $n=2$ with $x_H=-2$ (upper panel) and $x_H=-\frac{1}{2}$ (lower panel).}
\label{DM2-1}
\end{figure}

\begin{figure}[h]
\centering
\includegraphics[width=0.8\textwidth]{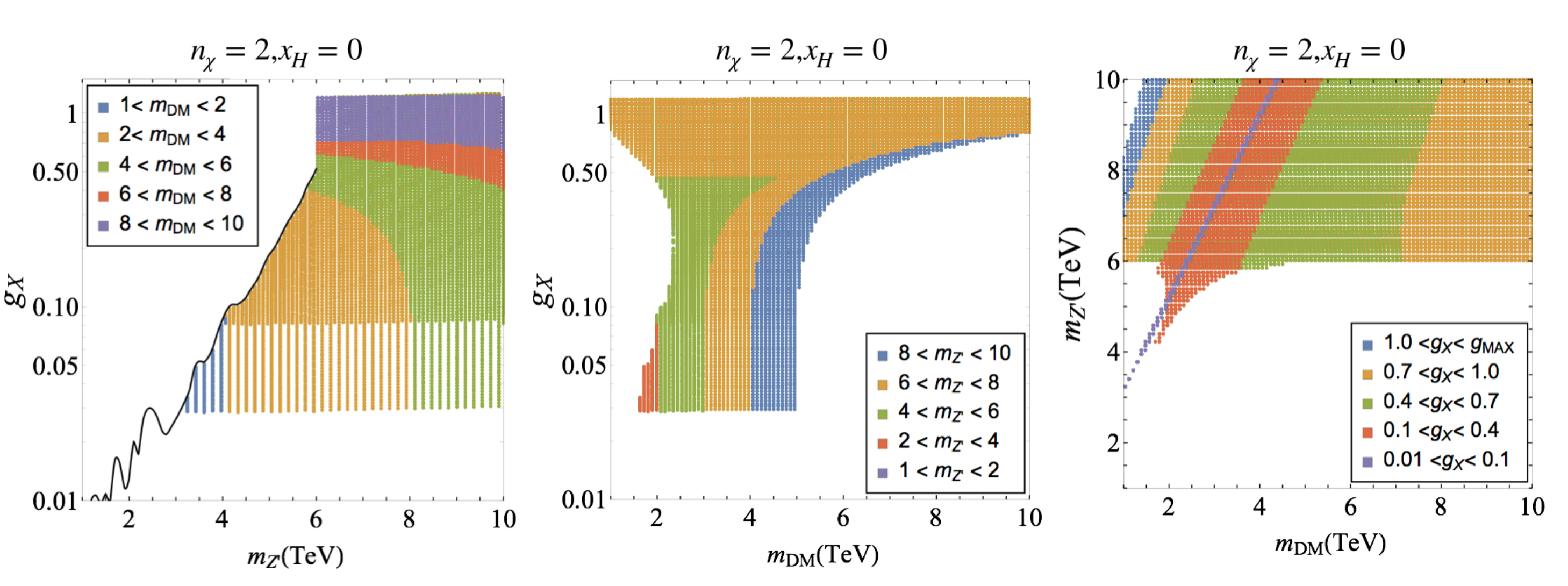} 
\includegraphics[width=0.8\textwidth]{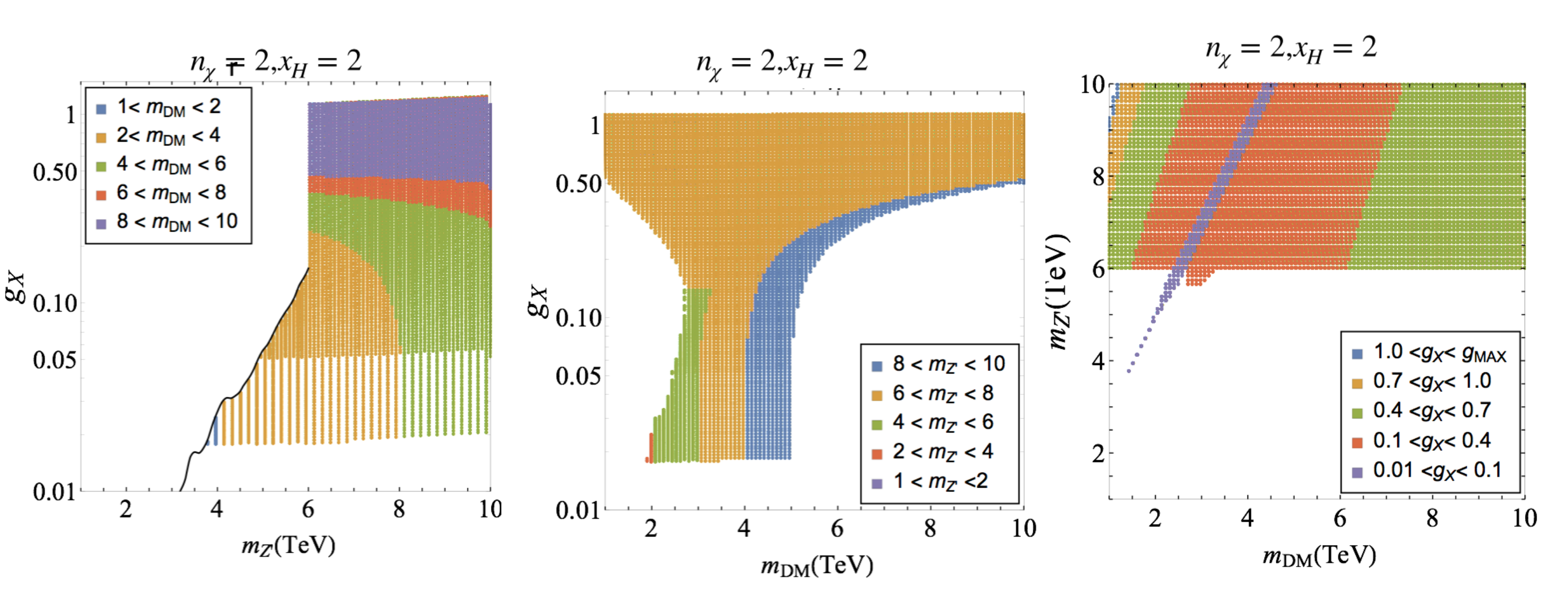} 
\caption{Same as Fig.~\ref{DM1-1} for $n=2$ with $x_H=0$ (upper panel) and $x_H=2$ (lower panel).}
\label{DM2-2}
\end{figure}

To study the bounds on $g_X$ and $M_Z^\prime$ we consider the collider limits as the strongest upper bounds. 
We scan over the parameters $g_X~\in~[0.01, 1.0]$, $m_{Z^\prime}~\in~[1~\rm{TeV}, 10~\rm{TeV}]$ and $m_{\rm{DM}}~\in~[1~\rm{TeV}, 10~\rm{TeV}]$ for $n_\chi=\frac{1}{3}$ and $n_\chi=2$. We find that the low $g_X$ region is less populated as the allowed region is restricted by the $Z^\prime$ pole. Fixing $n=\frac{1}{3}$ we find that with the increase in $x_H$ the parameter space for $m_{\rm{DM}}$ and $m_{Z^\prime}$ becomes thiner. In Fig.~\ref{DM1-1} we have plotted two benchmark scenarios for $x_H=-2$ (upper panel) and $x_H=-\frac{1}{2}$ (lower panel) with $n_\chi=\frac{1}{3}$, respectively. 
With the same $n_\chi$ we investigated the cases with $x_H=0$ (upper panel) and $x_H=2$ (lower panel) respectively in Fig.~\ref{DM1-2}. 
Similar nature of the parameters with the increase in $x_H$ can be found. Importantly we notice that for $x_H=-2$ the U$(1)_X$ charges of the SM quark and lepton doublets 
are zero which is a U$(1)_{\rm{R}}$ scenario. The U$(1)_X$ charge of the right handed up type quark for $x_H=-\frac{1}{2}$ is zero which do not contribute in the $\chi \chi \to \rm{SM}~\rm{SM}$ processes. The case with $x_H=0$ is the B$-$L scenario. The thermal averaged cross sections which are directly proportional to $x_H$ do not contribute in the B$-$L case. For $x_H=2$ all the SM particles contribute in the DM annihilation process. In this analysis we imposed the constraints from the DM relic abundance $\Omega h^2 \leq 0.12$. We study the case with $n_\chi=2$ for $x_H=-2$ (upper panel) and $x_H=-\frac{1}{2}$ (lower panel) in Fig.~\ref{DM2-1}. With the increase in $n_\chi$ we notice a significant enhancement in the parameter space which is contributed by the additional $Z^\prime Z^\prime$ mode.Similar behavior is observed in the case with $n_\chi=2$ for $x_H=0$ (upper panel) and $x_H=2$ (lower panel) in Fig.~\ref{DM2-2}.

%%%%%%%%%
%%%%%%%%%%%
%%%%%%%%%%%%%%%%%
%%%%%%%%%%%%%%%%%
%%%%%%%%%%%%%%
%\section{Conclusion}
%\label{conc}
%%%%%%%%%%%%%%%%%%%%

In this letter we study a model where the SM is extended by a general U$(1)_X$. 
This anomaly free model can explain the origin of the tiny neutrino mass by the seesaw mechanism.
The model includes an SM singlet scalar which can mix with the SM Higgs doublet and participate the phenomenological aspects of the model in a variety of ways. 
Such a model also responsible to study the origin of the DM which is not predicted by the SM and hence leads to a simple and suitable beyond the SM scenario. 
In this article we introduce a Dirac type DM candidate which is weakly interacting with the SM particles through a neutral beyond the SM gauge boson, $Z^\prime$.
Using the collider searches at the LHC we constrain the $U(1)_X$ coupling with respect to the $Z^\prime$ mass and apply the limits for the further studies.    
We calculate the thermal averaged cross sections of the leading processes followed by the DM relic abundance to find the allowed parameter space of the model. 
Hence applying the collider constraints and the constraints obtained by the DM relic abundance we represent the combined allowed regions of the model parameters.
%%%%%%%%%%
\section*{Acknowledgements}
\par{The works of A. D. and S. K. were supported in part by JSPS, Grant-in-Aid for Scientific Research, No. 18F18321. 
K. E. was supported in part by the Sasakawa Scientific Research Grant from The Japan Science Society.
S. K. was supported in part by Grant-in-Aid for Scientific Research on Innovative Areas, the Ministry of Education, Culture, Sports, Science and Technology, No. 16H06492 and also by JSPS, Grant-in-Aid for Scientific Research, Grant No. 18F18022, 20H00160.}
%%%%%%%%%%%%%%
\bibliography{bibliography}
\bibliographystyle{utphys}
\end{document}